\documentclass[aps,prb,showpacs,preprintnumbers,twocolumn]{revtex4-2}
\usepackage{amsmath,amssymb}
\usepackage{bm}
\usepackage{tipa}
\usepackage{upgreek}
\usepackage{comment}
\usepackage{mathrsfs}
\usepackage{graphicx}
\usepackage{braket}
\usepackage{mathbbol}
\usepackage{booktabs}
\usepackage{amssymb}
\usepackage{enumitem}
\usepackage{gensymb}
\usepackage{overpic}
\usepackage[normalem]{ulem}
\usepackage{color}
\usepackage[colorlinks,bookmarks=true,citecolor=blue,linkcolor=red,urlcolor=blue]{hyperref}
\usepackage{hyperref}

\usepackage{orcidlink}

\newcommand{\addACB}[1]{{\color{red}{#1}}}

\makeatletter
\let\old@makecaption=\@makecaption
\usepackage{subcaption}
\let\@makecaption=\old@makecaption
\makeatother

\usepackage{relsize}

\begin{document}

\author{Steven H. Simon}
\email{steven.simon@physics.ox.ac.uk}
\affiliation{Rudolf Peierls Centre, Oxford University, OX1 3NP, United Kingdom}

\author{Ajit C. Balram\orcidlink{0000-0002-8087-6015}}
\email{cb.ajit@gmail.com}
\affiliation{Institute of Mathematical Sciences, CIT Campus, Chennai 600113, India}
\affiliation{Homi Bhabha National Institute, Training School Complex, Anushaktinagar, Mumbai 400094, India}

\title{Phase Separation in the Putative Fractional Quantum Hall ${\cal A}$ phases.}

\begin{abstract}
We use several techniques to probe the wave functions proposed to describe the ${\cal A}$ phases by Das, Das, and Mandal [Phys. Rev. Lett. {\bf 131}, 056202 (2023); Phys. Rev. Lett. {\bf 132}, 106501 (2024); Phys. Rev. B {\bf 110}, L121303 (2024).]. As opposed to representing fractional quantum Hall liquids, we find these wave functions to describe states that clearly display strong phase separation. In the process of exploring these wave functions, we have also constructed several new methods for diagnosing phase separation and generating such wave functions numerically. Finally, we uncover a new property of entanglement spectra that can be used as a check for the accuracy of numerics. 
\end{abstract}

\maketitle

\section{Introduction}
Recently in Refs.~\cite{SSS1, SSS2, SSS3}, Das, Das, and Mandal (DDM) have studied what they call the ``${\cal A}$ phases" of electrons in Landau levels. These are meant to be a description of the ground states of a model Hamiltonian for the first excited Landau level which includes some approximation of Landau level mixing. After the publication of the first of these works one of us published a comment~\cite{SimonComment} claiming that their data was much more consistent with phase separation behavior rather than with a fractional quantum Hall (FQH) liquid. Despite the arguments presented in that comment, DDM replied~\cite{SSSReply} that they were unconvinced of clustering in the ${\cal A}$ phase. DDM then published further works along the same direction~\cite{SSS2, SSS3} proposing further wave functions --- which we believe also contain the same problems as those in their first work. The fact that these works were successfully published (i.e., were not rejected by referees) suggests that the message in the comment~\cite{SimonComment} was not sufficiently clear to the community as well as to the authors DDM. Indeed, in a one-page comment, it is difficult to give a complete discussion, and potentially the argument was not fully compelling at that point. The purpose of this paper is to give an extremely clear exposition of the physics that is observed in the ${\cal A}$ phase numerics. We will use several different techniques to give definitive and unambiguous evidence that the wave functions DDM are studying do not describe FQH liquids but rather describe phase separation (or ``clustering" more generally). Along the way, we create new tools that may be useful in other contexts. 

The outline of this paper is as follows. In section \ref{sec: numA} we briefly describe the numerics by DDM~\cite{SSS1, SSS2, SSS3} that led to the discussion of the ${\cal A}$ phase. In section \ref{sec: half} we focus on the half-filled Landau level case which was the case first examined by DDM. The principles we uncover here will tell most of the story. We argue that the wave functions they write down describe states that phase separate. One of the key points here is the effect of wave function antisymmetrization which we discuss in more detail in Appendix \ref{app: symmetrization}.  In section \ref{sub: diagnostic} we briefly describe a diagnostic for phase separation and apply it to the wave functions of DDM to show that they do indeed phase separate. More details about this diagnostic are given in Appendix \ref{app: cluster} along with its application to many other examples. In section \ref{sec: phasesep} we propose that we can reproduce the physics seen by DDM almost perfectly by starting with a phase-separated state and projecting it to zero angular momentum. In section \ref{sub: enta} we compare the entanglement spectrum of the wave function proposed by DDM to that of our trial wave function to show that they are essentially the same state.  In section \ref{sec: other} we extend the discussion to the wave functions proposed later by DDM in Refs.~\cite{SSS2, SSS3}. In section \ref{sec: toy} we describe a toy model Hamiltonian with a pure attractive interaction (first introduced by one of us in Ref.~\cite{SimonComment}) which is meant to demonstrate phase separation and we compare that to the results of DDM as well. We then give a brief conclusion in section~\ref{sec: conclusion}. In addition, in Appendix \ref{app: theorem} we state and prove a theorem about some of the properties that entanglement spectra must always have. This is quite useful for determining when numerics might be faulty (which we believe to be the case for DDM's work in Ref.~\cite{SSSReply}). 
 
\section{Numerics that suggested the ${\cal A}$ phases}
\label{sec: numA}

In FQH numerics, Landau level mixing is often neglected. However, in many experiments, the parameter that controls the strength of Landau level mixing (known as $\kappa$, the ratio of the Coulomb energy to the cyclotron energy) can be order unity. While it is usually not thought to have a strong effect, in some cases Landau level mixing can be crucial in deciding which of several possible states of matter might be realized~\cite{ Wojs10, SimonRezayi, Rezayi, Pakrouski, Mila}. 

One strategy for treating Landau level mixing is to treat $\kappa$ perturbatively at lowest~\cite{Bishara09, Peterson, Sodemann, SimonRezayiPerturbative} or next-to-lowest~\cite{Mila} order. While this approach has been extensively used in the past, one must be cautious not to use the approach out of its regime of validity. Unfortunately, it is often difficult to determine how large $\kappa$ can be before the neglected higher-order terms become important. While this approach is still valuable, it must be used with some caution. 

The work of DDM uses this perturbative approach to address Landau level mixing for filling fractions $2{<}\nu{<}3$. They find, fairly independent of the precise value of the filling fraction (!), that at some intermediate value of $0.9{\leq}\kappa{\leq}1.5$, there is a first-order transition into a phase which they call the ${\cal A}$-phase with a uniform, i.e., total orbital angular momentum $L{=}0$, ground state on the sphere. They have attempted to analyze this phase as if it were an FQH state. Further, they have found trial wave functions that have high overlaps with these ground states. As one of us suggested in a comment~\cite{SimonComment} and we will show in detail below, all of these wave functions describe phase-separated states and not FQH liquids. 

\section{The Half-Filled Landau Level Case}
\label{sec: half}

In the first paper by DDM~\cite{SSS1}, they proposed a wave function for $N$ (even) particles in $N_{\phi}{=}2N{-}1$ flux on the sphere~\cite{Haldane830}. To construct this wave function, take $N$ (even) particles and divide them into two groups of equal size (call them $A$ and $B$). We initially write a Halperin 113 wave function~\cite{Halperin} for the two species $A$ and $B$
\begin{eqnarray*}
 \Psi_{113} &=& \prod_{i < j;  i,j \in A} (u_i v_j - u_j v_i)  \prod_{n < m;  n, m  \in B}  (u_n v_m - u_m v_n)  \\ & & ~~~~~\times \prod_{p \in A, q \in B} (u_p v_q - u_q v_p)^3, 
\end{eqnarray*}
where $u_{r}{=}\cos(\theta_{r}/2)e^{i\phi_{r}/2}$ and $v_{r}{=}\sin(\theta_{r}/2)e^{{-}i\phi_{r}/2}$ are spinor coordinates of the $r^{\rm th}$ electron with $\theta_{r}$ and $\phi_{r}$ being its polar and azimuthal angles on the sphere. The wave function proposed by DDM is the antisymmetrization of $\Psi_{113}$ over all coordinates: 
\begin{equation}
\Psi_{\mathbb{A}[113]} = \mathbb{A}[ \Psi_{113} ]
    \label{eq: A113}
\end{equation}
where $\mathbb{A}$ is the anti-symmetrization operator. This can be interpreted as an alternating sum over all permutations of the particles, or equivalently one must sum over choosing the groups $A, B$ in all possible ways (again with appropriate signs corresponding to the parity of the number of exchanges). We will sometimes write $\mathbb{A}[113]$ as shorthand for this wave function. 

Let us start by examining the wave function $\Psi_{113}$ without antisymmetrization
It is known that this wave function phase separates --- the $A$ group goes to one side of the sphere and the $B$ group goes to the opposite side of the sphere~\cite{Gail}. This can be understood as follows. In the Laughlin plasma analogy~\cite{Laughlin83} all particles repel each other, however, the $A$ particles repel the $B$ particles three times as strongly as the $A$'s repel each other or the $B$'s repel each other. Thus, it is advantageous to move all the $A$'s as far away from the $B$'s as possible. An equivalent way of understanding this is to see that the amplitude of the wave function is maximized when all of the $A$'s are as far away from the $B$'s as possible (but keeping the $A$'s not quite on top of each other at the same time). 

\begin{figure}
    \centering
    \includegraphics[width=0.25\linewidth]{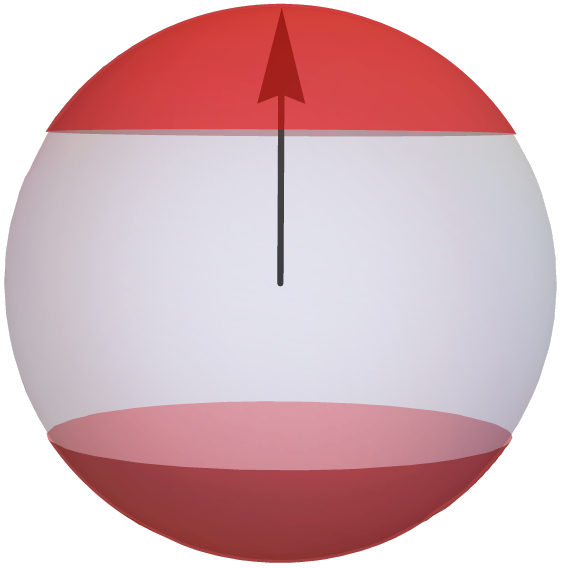}  \hspace*{1cm}  \begin{overpic}[abs,unit=1mm,scale=.25]{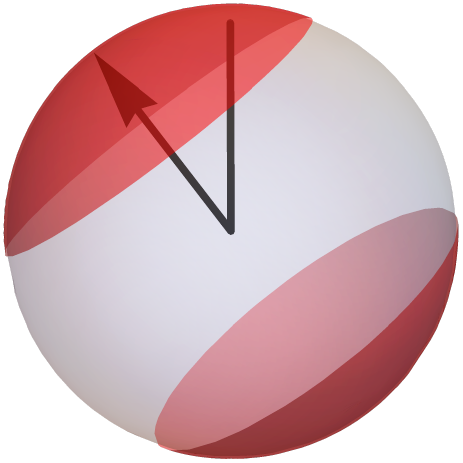}
\put(1,17){\color{blue}$\hat n$}
\end{overpic}  
    \caption{Sphere with two antipodal caps of electrons (depicted as red). The right picture has been rotated so that the caps are no longer on the north and south poles but are rather oriented at an angle $\hat n$}
    \label{fig: spheres}
\end{figure}

We thus conjecture we have a configuration that looks roughly like Fig.~\ref{fig: spheres} with two spherical caps of particles (of different species) on opposite (antipodal) sides of the sphere.

The wave function we are interested in, however, is not 113, but rather is the antisymmetrized version $\mathbb{A}[113]$. DDM argue~\cite{SSSReply, SSS3} that the antisymmetrization drastically changes the situation and prevents the system from separating in this way, resulting in a proper FQH liquid. While it is true that antisymmetrization can greatly change the properties of a wave function we claim it cannot do so in this particular case. Assuming the separation of the $A$ and $B$ species is very strong, then symmetrization or antisymmetrization can make nearly no difference. This is a fundamental principle of quantum mechanics: Unless two single-particle wave functions have a spatial overlap, it is not possible to distinguish with any local measurement whether the particles are distinguishable, bosonic (symmetrized wave functions), or fermionic (antisymmetrized wave function). If this principle is not familiar, we include Appendix \ref{app: symmetrization} explaining it in detail. 

\subsection{Phase-Separation Diagnostic}
\label{sub: diagnostic}

In Appendix \ref{app: cluster} we construct and test a simple diagnostic for determining when a wave function displays phase separation.  We emphasize that this diagnostic may be broadly useful for other systems, even outside of the quantum Hall world. While we refer the reader to the Appendix~\ref{app: cluster} for more details and examples [see Appendix~\ref{app: clusterexamples}], we will give a very short discussion here and show the data that clearly shows that the wave function $\mathbb{A}[113]$ is phase-separated. 
 
Our diagnostic is as follows. Let $R(k)$ be the average distance from an arbitrarily chosen particle to the $k^{\rm th}$ closest particle to that one. Roughly the distance to the $k^{\rm th}$ closest particle should satisfy $\pi [R(k)]^2  \rho {=} k$, where $\rho$ is the density of particles. We then construct 
\begin{equation}
    G(k) = [R(k+1)]^2 - [R(k)]^2.
\end{equation}
For a system where the particles are uncorrelated, $G(k)$ should be roughly constant $G {\approx} 1/(\pi \rho) {=} 2 N_\phi/N$ if we measure distance in units of magnetic length. However, a large peak in $G(k)$ for $k{=}k_0$ tells us that there is an anomalous increase in the distance to the $(k_0{+}1)^{\rm th}$ closest particle. In other words, if particles cluster into groups of $m$ particles that are tightly packed, but then there is a large distance between the clusters, then $G(k)$ has a large peak at $k{=}m{-}1$.

In Appendix \ref{app: clusterexamples} we explore several more wave functions to probe for phase separation or clustering.

\begin{figure}
    \centering
     \includegraphics[width=1\columnwidth]{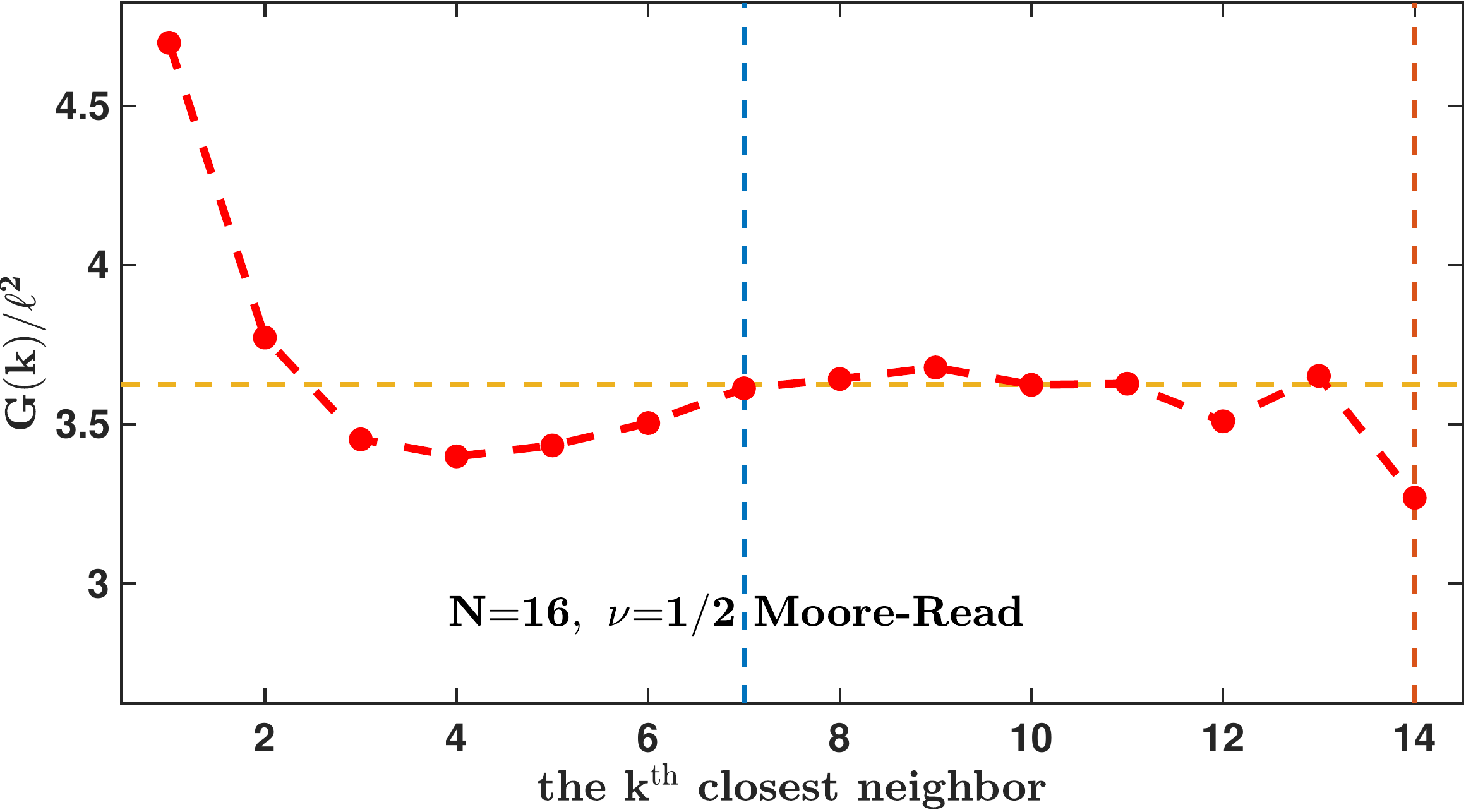}
     \phantom{.} \\
     \phantom{.}
      \includegraphics[width=1\columnwidth]{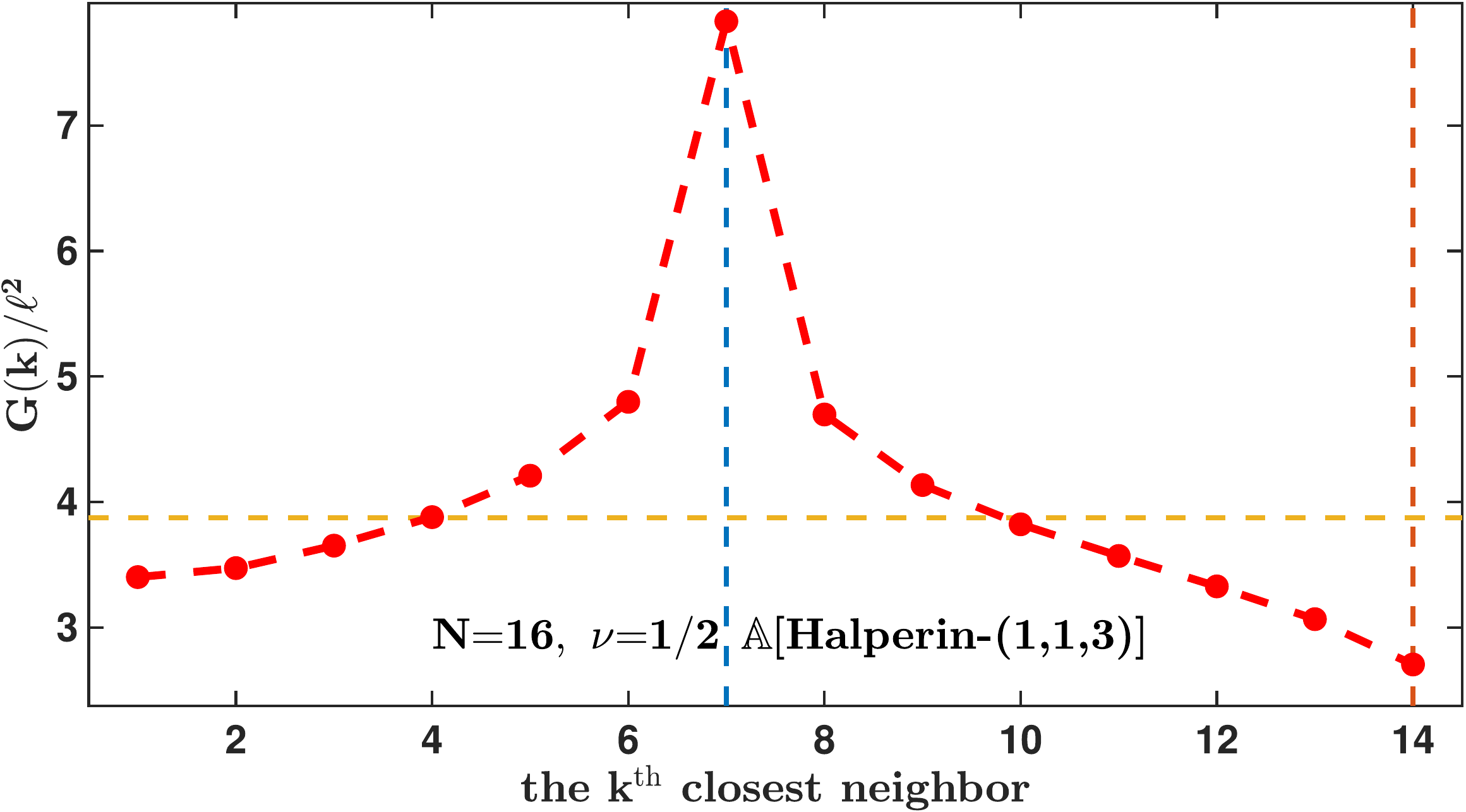}
    \caption{Two examples of the phase separation diagnostic function $G(k)$. In both plots, the yellow dashed line is $G{=}2 N_\phi/N$ which would result from an uncorrelated gas of particles. {\bf Top:}  Example of the Moore-Read wave function. This is chosen as a typical example of a FQH liquid. Here, there is some short-range structure for small $k$, but most of the data is very close to the yellow dashed line. {\bf Bottom:} The antisymmetrized Halperin 113 wave function. Here the huge peak at $k{=}7$ indicates that the system has broken up into two clusters of 8 particles that are far apart, exactly as depicted in Fig.~\ref{fig: spheres}. Note the difference in the vertical scales between the two plots. 
    }
    \label{fig:  MR_A113_N_16}
\end{figure}

Here, however, we will just show two examples in Fig.~\ref{fig:  MR_A113_N_16}. First, we show $G(k)$ for the Moore-Read state~\cite{Moore91} (occurs at $N_{\phi}{=}2N{-}3$), as a typical example of an FQH liquid -- which gives $G(k)$ as a very smooth function which, for $k{>}2$ is constant within about 10\% (pay attention to the vertical scale). All other FQH liquids we have tested are similar. Second, we show $G(k)$ for the Antisymmetrized Halperin 113 state --- which shows a large peak at $k{=}N/2{-}1$, clearly indicating phase separation of the type shown in Fig.~\ref{fig: spheres} with exactly $N/2$ particles per cluster.

\subsection{Comparison to Phase Separated Wave Function}
\label{sec: phasesep}

Here we are going to guess what the wave function should look like in the second quantized notation. 
We will assume that the system phase separates into antipodal caps as depicted in Fig.~\ref{fig: spheres}. Assuming the caps are at the north and south poles, the wave function is a simple Hartree-Fock state. The single particle orbital with the highest (lowest) angular momentum has $L_z {=}{ \pm} N_{\phi}/2$. Thus we can write the state with one cap of $N/2$ particles at the north pole and one cap of $N/2$ particles at the south pole as
\begin{equation}
 |\Psi(\hat z)\rangle = \prod_{m = -N_\phi/2}^{-N_\phi/2 +  N/2 - 1} c^\dagger_m    \prod_{n = N_\phi/2 - N/2 + 1}^{N_{\phi/2}} c^\dagger_n ~~| 0 \rangle. 
 \label{eq: psiz}
   \end{equation}
The notation $\hat z$ here is to note that the direction of the caps is towards the north and south poles as in the left of Fig.~\ref{fig: spheres}.

The wave function of Eq.~\eqref{eq: psiz} can be rotated to move the caps to any direction $\hat n$ (a unit vector) and ${-}\hat n$ as shown in the right of Fig.~\ref{fig: spheres}, by using a rotation operator $\hat{\mathcal{R}}(\hat n)$, i.e., 
$$
  |\Psi(\hat n)\rangle =  \hat{\mathcal{R}}(\hat n) |\Psi(\hat z)\rangle.
$$

The wave function  $|\Psi(\hat n)\rangle$ is not an angular momentum eigenstate. However, if we integrate over all directions $\hat n$ we obtain a rotationally invariant result,  which should therefore be an $L{=}0$ eigenstate, i.e., 
\begin{equation}
 |\Psi_{L=0}\rangle  =  \int d\hat n   \,\,
\hat{\mathcal{R}}(\hat n) \,\, |\Psi(\hat z)\rangle    .
\label{eq: rotate}
\end{equation}
Some (complicated) applications of orbital angular momentum $\hat L_{x,y,z}$ operators can implement this integration over all directions. However, the result must act linearly on $|\Psi(\hat z)\rangle$ and must give a result that is an $L{=}L^2{=}0$ eigenstate. The only possibility is that we are implementing (up to a normalization constant) 
\begin{equation}
|\Psi_{L=0}\rangle = \mathcal{P}_{L=0} |\Psi(\hat z)\rangle,
\label{eq: ourtrial}
\end{equation}
where $\mathcal{P}_{L=0}$ is the projection operator to the space of states with zero orbital angular momentum. It is fairly easy to construct the wave functions in Eq.~\eqref{eq: ourtrial} numerically for small systems using the methods outlined in Refs.~\cite{Sreejith13, Balram20a}. We can then compare the phase-separated wave function Eq.~\eqref{eq: ourtrial} with the antisymmetrized Halperin 113 state, $\mathbb{A}[113]$, by computing the overlaps between the two (shown in Table~\ref{tab: overlaps}). The very high overlap between the two states, even with a Hilbert space dimension nearing 1000, is impressive and should put to rest any doubts as to the physics of this state. The fact that the overlap is not perfect tells us that the $\mathbb{A}[113]$ state does not form perfect (fully packed) caps, but rather allows some tiny fluctuations.

\begin{table}[]
    \centering
    \begin{tabular}{c|c|c}
        $N$  &   dim($L^2{=}0$) &   $~~~~|\langle \,\Psi_{L{=}0} \, | \, \mathbb{A}[113] \rangle \,|^2$   \\
        \hline
6  &  3  &      0.9147  \\ 
8  &  7  &     0.8966 \\ 
10 &  24  &     0.8916 \\ 
12  & 127   &    0.8877 \\
14  & 802  &   0.8849
    \end{tabular}
    \caption{Overlap between the phase separated state $|\Psi_{L{=}0}\rangle$ defined in Eq.~\eqref{eq: ourtrial}  and the antisymmetrized Halperin 113 state, $\mathbb{A}[113]$, defined in Eq.~\eqref{eq: A113}, as a function of the number of electrons $N$ at flux $N_\phi{=}2N{-}1$. The middle column is the total dimension of the $L{=}0$ Hilbert space.}
    \label{tab: overlaps}
\end{table}

\subsection{Entanglement Spectra}
\label{sub: enta}

\begin{figure}
    \centering    \includegraphics[width=\linewidth]{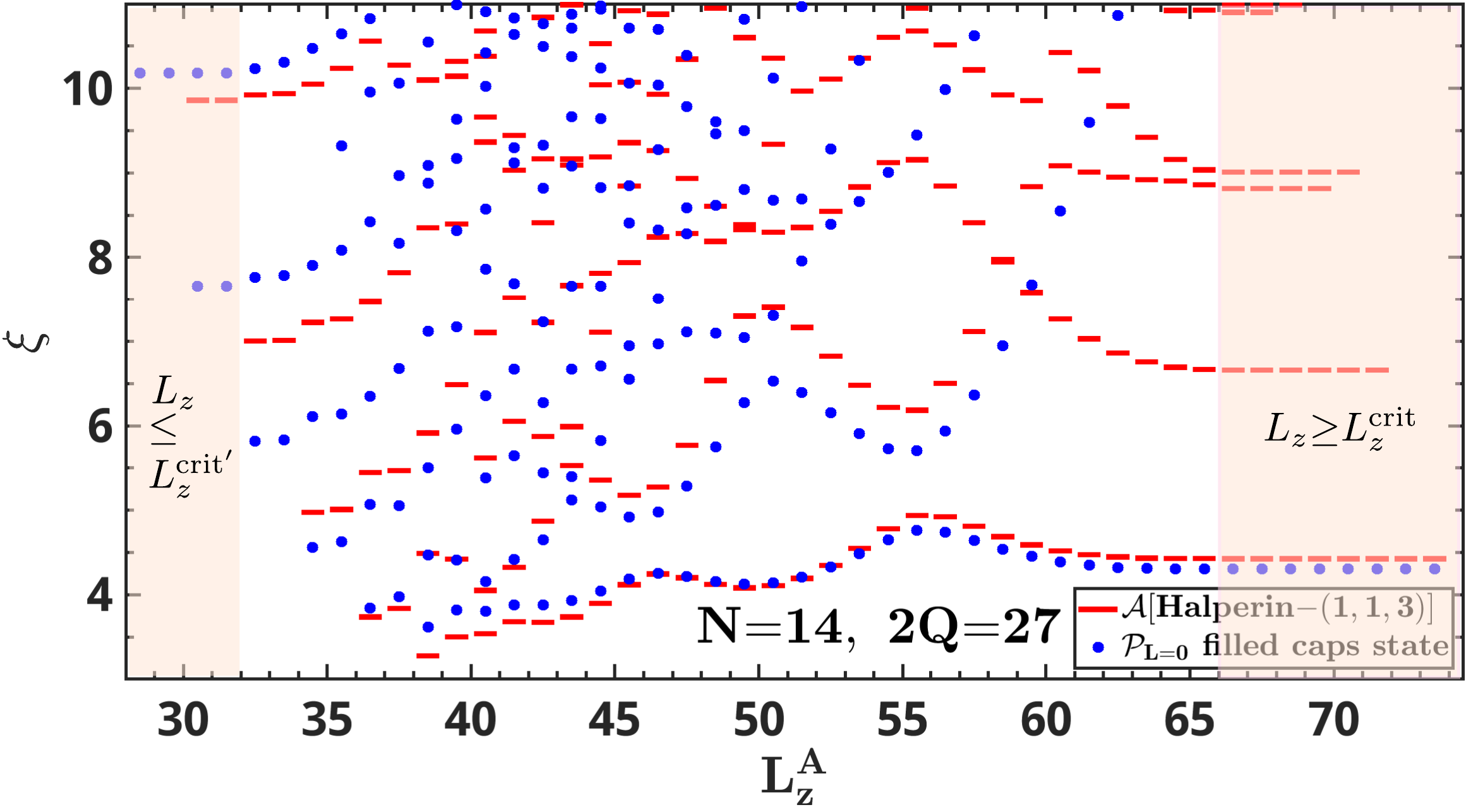}
    \caption{Comparison of the entanglement spectra of $A[113]$ versus the $|\Psi_{L{=}0}\rangle$. Here $N_\phi{=}27$ and the system is partitioned into $7{+}7$ electrons at the equator. The different branches in the entanglement spectra are flat for $L_z{\geq}L_z^{\rm crit}$ and $L_z{\leq}L_z^{\rm crit'}$ ($L_z^{\rm crit}$ and $L_z^{\rm crit'}$ are defined in Appendix \ref{app: theorem}).}
    \label{fig: entanglementnew}
\end{figure}

One of the features of note about the data presented by DDM in Ref.~\cite{SSS1} is that the entanglement spectra do not look anything like spectra of other FQH states~\cite{RegnaultReview}. In particular, there is a branch of the entanglement spectrum that stays at low ``energy" out to very high angular momentum (see Fig.~\ref{fig: entanglementnew}). This is a clear alert that we are not examining a liquid. 

If we split the sphere in half with the restriction that $N/2$ electrons are in either half, one can obtain a maximum angular momentum in the northern hemisphere given by~\footnote{This counting alerted one of the present authors of an error in the caption of Fig. 3c in the arxiv version of Ref.~\cite{SSS1}. The present author alerted the authors of Ref.~\cite{SSS1} to the error, which was fixed in the final published version.} 
\begin{equation}
 L_z^{\rm max} = \sum_{m = N_{\phi}/2 - N/2 + 1}^{N_{\phi}/2} m = \frac{1}{8} N [2 (N_\phi + 1) - N]      \label{eq: Lzmax}
\end{equation}
where the electrons fill the cap at the north pole. For a correlated liquid state, this clustering of electrons is exceedingly unlikely (Indeed, for the exact Laughlin state~\cite{Laughlin83} at $\nu{=}1/p$ with $p{>}1$ the amplitude of this Fock state is zero.). This means that the entanglement ``energy" for that angular momentum must be high (or infinite in the case of Laughlin). However, for the phase-separated states, as described in the prior subsection [see Eq.~\eqref{eq: ourtrial}], this Fock state is clearly very prominent, and hence the entanglement energy is very low. Further, by considering the piece of the wave function where the cap has been rotated away from the north pole in Eq.~\eqref{eq: rotate}, we see that this branch of the entanglement energy should remain at low energy as $L_z$  of the northern hemisphere is decreased until part of the cap of filled states crosses the equator (see also the discussion in Appendix~\ref{app: theorem}). 

A comparison of the entanglement spectra of $\mathbb{A}[113]$ versus the $|\Psi_{L{=}0}\rangle$ is shown in Fig.~\ref{fig: entanglementnew}. The two states clearly show the same physics at low entanglement energy, and some of the features even match fairly well at higher entanglement energies. 

We take this opportunity to mention a general theorem about entanglement spectra which we prove in Appendix~\ref{app: theorem}. For any 
orbital entanglement spectra~\cite{Li08} of an $L{=}0$ state in a Landau level on a sphere (with $N$ even, $N_\phi$ odd)  partitioned at the equator with $N/2$ particles on each side, we define $L_z^{\rm crit} {=} L_z^{\rm max} {-} (N_\phi {-} N {+} 1)/2$. The theorem states that all branches of the entanglement spectrum must be exactly flat for $L_z {\geq} L_z^{\rm crit}$. In Fig.~\ref{fig: entanglementnew}, $L_z^{\rm max} {=} 73.5$ and $L_z^{\rm crit} {=} 66.5$. We see that all modes above $L_z^{\rm crit}$ are indeed flat (to 8-digit precision in our numerics) as required. 
Similarly, the lowest possible angular momentum mode should be at $L_z^{\rm min}{=}N^2/8$, and modes should be exactly flat up to $L_z^{\rm crit'}{=}L_z^{\rm min}{+}N/2$. In this case we have $L_z^{\rm min}{=}24.5$ and $L_z^{\rm crit'}{=}31.5$. Although many of the relevant data points are off the top of the plot, we do find this data to agree with our theorem.

\section{Other Fillings}
\label{sec: other}

One of the notable things about the data shown in Refs.~\cite{SSS1, SSS2, SSS3} is that the entanglement spectra seem to have the same structure {\it independent} of the precise filling fraction we are examining (so long as we have an even number of electrons in the system and we examine only the $N/2{+}N/2$ sector). This is certainly not a property of any known FQH liquid. However, once we realize that we are looking at phase-separated states as described in section \ref{sec: phasesep} this observation becomes very natural. Indeed, any system that phase separates into two equal clusters that repel each other will have essentially the same properties. 

Independent of the precise value of $N$ (chosen even) or $N_\phi$, one can construct a wave function with antipodal caps as in Eq.~\eqref{eq: psiz} and then one can project to zero total orbital angular momentum as in Eq.~\eqref{eq: ourtrial}. Such a phase-separated wave function matches all of the properties of all of the data displayed in Refs.~\cite{SSS1, SSS2, SSS3}. In particular, all of the entanglement spectra have a low energy branch that goes all the way out to $L_z^{\rm max}$ as defined by Eq.~\eqref{eq: Lzmax} indicating the presence of a maximally compressed cap at the north pole. 

In Refs.~\cite{SSS1, SSS2, SSS3} many wave functions are written down which all display good overlaps with the considered trial states at a variety of filling fractions. It is worth examining these wave functions and figuring out why these seem to work well.

\subsection{Useful Notation for Jastrow Factors}
\label{sec: jastrowsection}

We will consider systems in the spherical geometry~\cite{Haldane830}. Using spinor coordinates $(u,v)$ for a particle, our wave functions will be made up of Jastrow factors between particle $i$ and $j$
$$
\tilde  z_{ij} = (u_i v_j - v_i u_j)
$$
In our wave functions, we will divide the particles into various groups. For a group of particles $G$, we write a Jastrow factor between all particles in this group as
$$
J_{GG} =  \prod_{i < j;  i,j \in G} \tilde z_{ij}
$$
and for two different groups $G$ and $H$, we write a Jastrow factor between groups as
$$
   J_{GH} = \prod_{i \in G, j \in H}  \tilde z_{ij} ~~~~~~ G \neq H
$$
We also write the full Jastrow factor as 
\begin{equation}
 J = \prod_{i < j} \tilde z_{ij}
\end{equation}
for all particles in the system. 

As an example, the Halperin 113 wave function can be written as $J_{AA}^1 J_{BB}^1 J_{AB}^3{=}J J_{AB}^2$.

\subsection{Wave Functions of Ref.~\cite{SSS2}}
\label{sub: generalwf}

In Ref.~\cite{SSS2}, wave functions meant to describe FQH effect at $\nu{=}n/(nm{-}1)$ are constructed by dividing the particles into $2n$ groups of equal size which we label $A_1, {\ldots}, A_n$ and $B_1, {\ldots}, B_n$. The generalized wave function is given by
$$
\Psi(n,m) = \mathbb{A}\left[ J \prod_{k=1}^n J_{A_k B_k}^{2(m-2)}  \prod_{i,j=1; i\neq j}^n J_{A_i B_j}^{2 (m-1)} \right]
$$
Neglecting the leading Jastrow factor $J$ and neglecting the antisymmetrization out front, the amplitude of this wave function is maximized by putting all of the $A$ particles antipodal to all of the $B$ particles. Using the plasma analogy to be more precise, we have a model where (from the leading Jastrow factor) all particles repel each other, but due to the $J_{AB}$ factors, the $A$ particles repel $B$ particles at least $2(m{-}2)$ times more strongly [and sometimes $2(m{-}1)$ times more strongly] than any $A$ particle repels any other $A$ particle. For $m{>}2$, this means that the ground state will be phase separated with $A$ particles as far away as possible from the $B$ particles --- forming antipodal caps as in Fig.~\ref{fig: spheres}. 

Again, we then must worry about the antisymmetrization. The wave function $\Psi(n,m)$ is already properly antisymmetric within all of the $A$ groups and also within all of the $B$ groups. The only thing one needs to do is antisymmetrize which particles are $A$'s and which particles are $B$'s, i.e., antisymmetrize between the two separated antipodal caps. However, as discussed for the 113 case above in section \ref{sec: half} and argued in more detail in Appendix~\ref{app: symmetrization}, symmetrization or antisymmetrization between spatially separated regions makes little difference. 

In Appendix \ref{app: clusterexamples} in Fig.~\ref{fig:psi23} we show the cluster diagnostic $G(k)$ for the trial wave function $\Psi(2,3)$ which is meant to describe $\nu{=}2/5$. The large peak in $G(k {=} N/2 {-}1)$ shows that the system has formed two antipodal caps exactly analogous to the case of $\mathbb{A}[113]$ that we discussed above.

\subsection{Wave Functions of Ref.~\cite{SSS3}}

In Ref. \cite{SSS2}, a wave function meant to describe $\nu{=}6/13$ is constructed. Here we divide the particles into four groups, $A, B, a, b$ such that the number of particles in $A$ is the same as the number in $B$ and similarly the number in $a$ is the same as the number in $b$, but $A$ has twice the number as $a$ and $B$ has twice the number as $b$. We can then write the wave function as 
$$
 \Psi_{6/13} = \mathbb{A}\left[ J  J_{AB}^2 J_{Ab}^3 J_{aB}^3 J_{ab}^1 \right] 
$$
We again start by ignoring the antisymmetrization. 
If we put the $A$'s and $a$'s in one group we call $\alpha$ and we put the $B$'s and $b$'s in a group we call $\beta$, neglecting the 
prefactor of $J$, we see that the $\alpha$'s all repel the $\beta$'s but the $\alpha$'s do not repel other $\alpha$'s. Putting the $J$ factor back in and using the plasma analogy, we find that any $\alpha$ particle will repel any $\beta$ particle at least twice as strongly as any $\alpha$ repels any other $\alpha$ (and up to four times as strongly). We conclude that phase separation where the $\alpha$'s cluster antipodally to the $\beta$'s as in Fig.~\ref{fig: spheres} is likely (this can be checked easily with Monte Carlo). 

Here the antisymmetrization now looks a bit less trivial. Within the $\alpha$ group, the wave function is antisymmetrized within the $A$'s and also within the $a$'s but it is not antisymmetrized between the $A$'s and the $a$'s. However, we note that if the $\alpha$'s are far away from the $\beta$'s, then the precise position of any particle does not matter much. For example, in considering the wave function for any $\alpha$ particle $i$ we can approximately write a wave function for this particle as
\begin{equation}
 (u_i V - U v_i)^\gamma 
 \label{eq: mean}
\end{equation}
where $(U, V)$ is the spinor coordinate of the center of mass of the antipodal $\beta$-cap. Here, the exponent of $\gamma {=} 5 N/3$ comes from counting up all of the powers in the wave function between $i$ and all of the $\beta$ particles and it is independent of whether we chose $i$ as an $A$ particle or an $a$ particle. One must be careful to keep track of the fact that the $B$ group has $N/3$ particles whereas the $b$ group has $N/6$. For example, if particle $i$ is an $A$ particle then it accumulates $N/2$ powers from the factor $J$ (with only the $N/2{=}N/3{+} N/6$ particles in $B,b$ contributing) plus $2(N/3)$ powers from the $J_{AB}^2$ factor, plus $3(N/6)$ powers from $J_{Ab}^3$ resulting in $N (1/2{+}2/3{+}3/6){=}5 N/3$. The exponent is the same if particle $i$ had been an $a$ particle. 

Thus, within this level of approximation, the antisymmetrization within each of the groups $\alpha$ and $\beta$ does not matter much, and we then only need to antisymmetrize between the two caps, which, as we have already argued, also does not matter much. The fact that this wave function has a large entanglement weight (low entanglement energy) at $L_z^{\rm max}$ confirms the phase-separated structure. 

The proposed wave function for $\nu{=}5/13$ is similar. Here we divide the particles into six groups which we label $A_1, A_2, B_1, B_2, a, b$ where the number of particles in the capitalized groups is twice that in the small letter groups. The wave function is then of the form:
$$
 \Psi_{5/13} = \mathbb{A}\left[ J  J_{A_1 B_1}^2 J_{A_2 B_2}^2 J_{A_1 B_2}^4 J_{A_2 B_1}^4 J_{A_1 b}^4 J_{A_2 b}^4 J_{B_1 a}^4 J_{B_2 a}^4\right] 
$$
Again we start by ignoring the antisymmetrization. We group $A1,A2,a$ into group $\alpha$ and $B_1, B_2, b$ into group $\beta$. Ignoring the leading factor of $J$ none of the $\alpha$'s repel other $\alpha$'s and similarly, none of the $\beta$'s repel other $\beta$'s, but $\alpha$'s do repel $\beta$'s suggesting that they will form antipodal caps. Again we have the apparent issue that the wave function is not anti-symmetric between all of the particles within $\alpha$ (nor within $\beta$). However, if the $\alpha$'s and $\beta$'s form antipodal caps we can write a wave function for any $\alpha$ particle as Eq.~\eqref{eq: mean} with a value $\gamma {=} 21N/10$, in which case the full antisymmetrization can be ignored as above. Again, the proof that this works comes from the observation of the form of the entanglement spectrum.

\subsection{Further Details of Entanglement Spectrum}

In Ref.~\cite{SSS3} entanglement spectra are also shown for certain unequal partitions --- splitting the sphere at the equator but demanding that $N_1$ and $N_2$ electrons are on the two sides with $N_1{<}N_2$. In this case, the low energy branch extends to $L_z^{\rm max}$ given by the formula of Eq.~\eqref{eq: Lzmax}, but with $N/2$ replaced by $ N_1$ thus corresponding to a cap of $N_1$ particles at the north pole (this gives $L_z^{\rm max}{=}44$ for Fig. 3b  of Ref.~\cite{SSS3} and $L_z^{\rm max}{=}34.5$ for Fig. 3d of Ref.~\cite{SSS3} in agreement with their data). Unlike the case where we partition the sphere into $N/2{+}N/2$ in the two hemispheres, in these spectra, the entanglement weight increases strongly as $L$ approaches $L_z^{\rm max}$. This feature is entirely expected. To obtain this value of $L_z^{\rm max}$ we must have a configuration with $N_1{<}N/2$ at the north pole and no other electrons in the northern hemisphere. Such a configuration is of very small amplitude in the ground state which is built from antipodal caps having exactly $N/2$ electrons each.

    \begin{figure*}
    \centering
     \includegraphics[width=2\columnwidth]{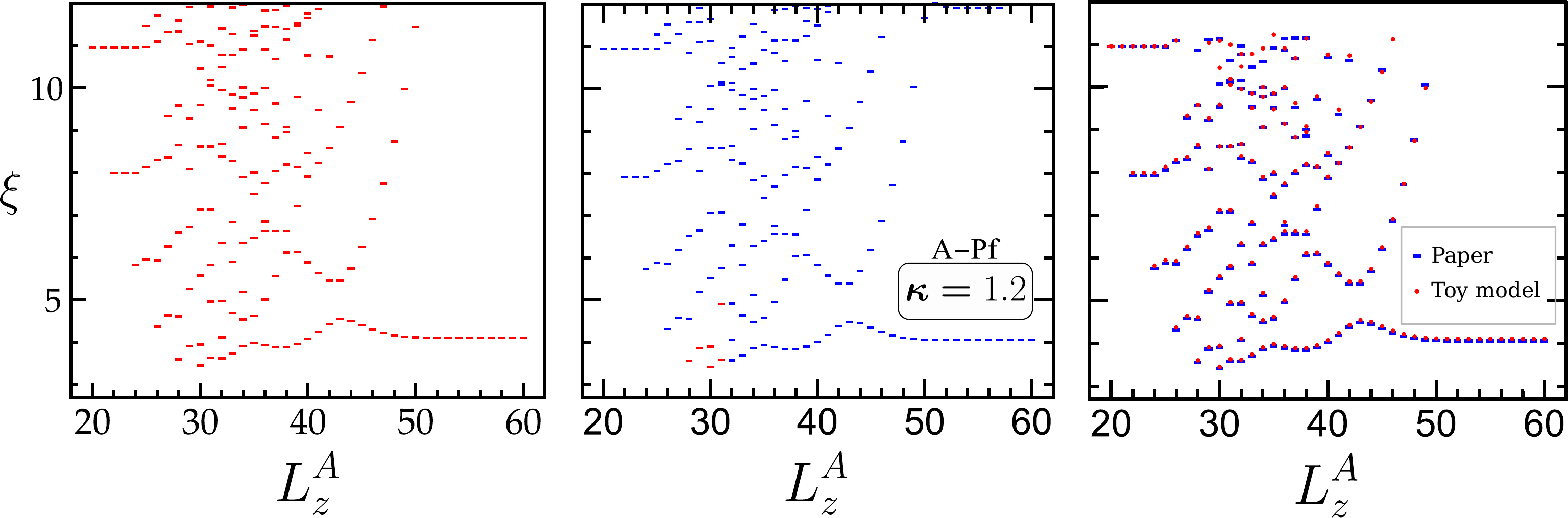}
    \caption{
Comparison of the entanglement spectrum of the state found by DDM~\cite{SSS1} (middle) and the state generated as the lowest energy $L=0$ state of the toy-model short-ranged attractive Hamiltonian [Eq.~\eqref{eq: toymodel}] on the left. The middle is an overlay of the two other panels. This is $N{=}12$ electrons and $N_\phi{=}25$ flux. This figure is reproduced from Ref.~\cite{SimonComment} having been provided by G. J. Sreejith.}
    \label{fig: entanglementfromcomment}
\end{figure*}    

\section{Comparison to Attractive Toy-Model Hamiltonian }
\label{sec: toy}

To demonstrate that the data in Ref.~\cite{SSS1} is phase separated as shown in Fig.~\ref{fig: spheres}, in the comment~\cite{SimonComment}, one of us considered a toy model of pure short-range attraction. This model can be described as a negative $V_1$ Haldane pseudopotential~\cite{Haldane830}, or equivalently~\cite{Trugman85}
\begin{equation}
V({\bf r}) = -\nabla^2 \delta({\bf r})     ~~~~ \mbox{Toy Model}  \label{eq: toymodel}
\end{equation}
The absolute ground state of this Hamiltonian should be a single cluster of electrons. This forms a maximum $L$ state, which does not match the wave functions considered by DDM. However, when we consider the lowest lying $L{=}0$  ($L^2{=}0$) state, we find an almost perfect agreement as shown in the comment~\cite{SimonComment} and reproduced in Fig.~\ref{fig: entanglementfromcomment} here. It is then worth considering what the lowest lying $L{=}0$ state looks like. Obviously, such a state should be as clustered as possible, while still being overall rotationally invariant ($L{=}0$). Averaging over rotations [analogous to Eq.~\eqref{eq: rotate}] will annihilate states that have no $L{=}L_z{=}0$ component. The simplest way to do this (and the most energy-efficient way) is to split the cluster into two antipodal caps and average over directions (or equivalently project to $L{=}0$) as we have done. DDM argued in Ref.~\cite{SSSReply} that a state that is not the overall ground state could be very different from the overall ground state. However, in the case of Eq.~\eqref{eq: toymodel}, it is quite clear that strong clustering must be favored for any reasonably low-lying state.

Indeed, looking at the effective Hamiltonian of DDM (see footnote 40 of Ref.~\cite{SSS1}), we see that as the Landau-level mixing parameter $\kappa$ is increased, the corrections to the (particularly short-ranged) interaction become increasingly attractive and will overwhelm the bare interaction. One might expect exactly the same physics as our toy model Hamiltonian. However, at least for some values of $\kappa$, DDM found $L{=}0$ to be the absolute ground state, unlike our toy model. However, we must also remember that there is a long-range piece of the Coulomb interaction that is not in our toy model, which remains repulsive even for large $\kappa$  (i.e., the Landau-level mixing correction remains very small for the long-ranged part). Thus, we have to balance the short-range attraction that favors clustering with a long-range repulsion that prevents the large build-up of charge. Indeed, this balance is familiar in many other contexts. For example, in high Landau levels, it is known that electrons cluster, but into clusters of finite size~\cite{Moessner, Fogler, Koulakov, Fogler2001}. In this particular regime, it appears that clustering into exactly two clusters that then repel is favored. One might expect that for larger system sizes, the system breaks up into a larger number of clusters to preserve this balance, but none of the data presented by DDM seem to indicate this (The data presented in Fig. 2a of Ref~\cite{SSSReply} might look as if it behaves differently from the other data at larger system size, but as discussed in Appendix \ref{sub: error} we believe this data to be unreliable.).

We note an important subtlety in the use of the toy model given in Eq.~\eqref{eq: toymodel}. Within a single Landau level, any two-body interaction is particle-hole symmetric. This means that (at least on an infinite plane) it would be equally valid to say that the holes form clusters as to say that the electrons form clusters. On the sphere, with a two-body interaction, the situation is perfectly particle-hole symmetric only when $N_{\rm orb}{=}N_{\phi}{+}1{=}2N$. In this case, the two states of clustered antipodal electrons and clustered antipodal holes are degenerate (and inequivalent).   However, for $N_{\phi}{>}2N{-}1$ the clustering of electrons is lower energy while for $N_{\phi}{<}2N{-}1$ the clustering of holes is lower energy (The data in Fig.~\ref{fig: entanglementfromcomment} is in the electron clustering regime.). In contrast, the Hamiltonian used by DDM (see footnote 40 of Ref.~\cite{SSS1}) contains three-body terms (of negative sign) which strongly break the particle-hole symmetry and favor electron clustering at all fillings.

\section{Conclusions}
\label{sec: conclusion}

Several interesting questions remain about these systems. Considering the simplicity of the state written in the form of Eq.~\eqref{eq: ourtrial} it seems that one might be able to calculate the full entanglement spectrum analytically, or at least approximate it. We have understood why the low entanglement energy mode extends out to $L_z^{\rm max}$ and why it starts flat as we decrease $L_z$. However, there is a clear structure at lower $L_z$ which would be nice to understand more completely. For $L_z$ just slightly less than $L_z^{\rm crit}$ or slightly greater than $L_z^{\rm crit'}$ (see Appendix \ref{app: theorem} for the definition of $L_z^{\rm crit}$ and $L_z^{\rm crit'}$) it seems that the branches are {\it almost} flat. This can likely be understood in detail by realizing that (for $L_z$ near $L_z^{\rm crit}$ for example) as we lower the angular momentum of a branch using $L_-$ (as in Appendix \ref{app: theorem}) when we are just slightly below $L_z^{\rm crit}$ only a very few states of the Hilbert space have any electrons that have crossed through the equator, so the theorem about flatness {\it almost} works.  Other features would also be interesting to understand. In particular, it would be interesting to understand the low $L_z$ end of the lowest energy branch which attracted the interest of DDM~\cite{SSS1, SSS2, SSS3}.

Given the number of avenues we have used to examine the properties of the ${\cal A}$ phase, there should be no further debate about the nature of these states. Indeed, one might even wonder why we put so much work into this issue. Although these states are not FQH liquids, they are interesting nonetheless. Further, the new tools we have used to probe these states may be broadly useful in other contexts. So the exercise has certainly not been without rewards. 

\vspace*{10pt}

\begin{acknowledgments}
The work was made possible by financial support from the Science and Engineering Research Board (SERB) of the Department of Science and Technology (DST) via the Mathematical Research Impact Centric Support (MATRICS) Grant No. MTR/2023/000002. Computational portions of this research work were conducted using the Nandadevi and Kamet supercomputers maintained and supported by the Institute of Mathematical Science's High-Performance Computing Center. Some numerical calculations were performed using the DiagHam package~\cite{diagham}, for which we are grateful to its authors. S.H.S. acknowledges support from EPSRC grant EP/X030881/1. We thank G. J. Sreejith for providing Fig.~\ref{fig: entanglementfromcomment} and for useful discussions. 
\end{acknowledgments}

\appendix

\section{(Anti)-Symmetrization}
\label{app: symmetrization}

Here we examine the issue of (anti)symmetrization. As argued by DDM~\cite{SSSReply, SSS2} (anti)symmetrization of a wave function can strongly change its properties—for example, the Abelian 331 state when anti-symmetrized gives the non-Abelian Moore-Read state. 

However, in Ref.~\cite{SimonComment} one of us argued that for a phase-separated wave function, the (anti)symmetrization will do almost nothing. Indeed, we have found that plotting $G(k)$ for the 113 and the antisymmetrized 113 states are almost numerically identical. There is a very good reason for this. For symmetrization between two wave functions to have any effect on any measurable quantity, the two wave functions must overlap. Thus if we have two clusters that have strongly repelled each other,  and the overlaps between the clusters are close to zero, then no locally measurable quantity can distinguish the symmetrized and unsymmetrized wave functions. 

To see this principle in more detail, let us consider two normalized wave functions $\phi_1(r)$ and $\phi_2(r)$ such that there is no point $r$ where both $\phi_1(r){\neq}0$ and $\phi_2(r){\neq}0$. Suppose we construct the (unsymmetrized) wave function $\phi_1(r_1) \phi_2(r_2)$ and we want to measure some operator $\hat O(r_1, r_2)$ which is symmetric between $r_1$ and $r_2$. For example, we can measure the total density or distance between the two particles. Any operator that we can write in the second quantized notation (which preserves particle number) is necessarily symmetric between particles. 
We write
\begin{equation}
\label{eq:  Oexp}
 \langle O \rangle = \int dr_1 dr_2  \,\,  \phi_1^*(r_1) \phi_2^*(r_2) \hat O(r_1, r_2) \phi_1(r_1) \phi_2(r_2).
\end{equation}
Now let us consider an antisymmmetrized wave function. We have
$$
 \frac{1}{\sqrt{2}}\left[ \phi_1(r_1) \phi_2(r_2) - \phi_1(r_2) \phi_2(r_1)  \right].
$$
We thus calculate
\begin{widetext}
\begin{equation} \nonumber
    \langle O \rangle = \frac{1}{2} \int dr_1 dr_2    \left[ \phi_1^*(r_1) \phi_2^*(r_2) - \phi_1^*(r_2) \phi_2^*(r_1)  \right]
 \hat O(r_1, r_2)\left[ \phi_1(r_1) \phi_2(r_2) - \phi_1(r_2) \phi_2(r_1)  \right] \nonumber.
\end{equation}
We multiply out the four terms here, but we discover that the two cross terms vanish. Cross factors like  $\phi_1^*(r_1) \phi_2(r_1)$ are identically zero since there is no overlap between $\phi_1(r_1)$ and $\phi_2(r_1)$ so there is no value of $r_1$ which will make this nonzero. Thus we end up with 
\begin{equation}
 \langle O \rangle = \frac{1}{2} \int dr_1 dr_2  \,\,  [ \phi_1^*(r_1) \phi_2^*(r_2) \hat O(r_1, r_2) \phi_1(r_1) \phi_2(r_2) +  \phi_2^*(r_1) \phi_1^*(r_2) \hat O(r_1, r_2) \phi_2(r_1) \phi_1(r_2) ]    \label{eq:  Oexp2}.
\end{equation}
\end{widetext}
Since the operator $\hat O$ is assumed symmetric in its argument, the two terms in Eq.~\eqref{eq:  Oexp2} are identical. Thus Eq.~\eqref{eq:  Oexp2} and Eq.~\eqref{eq:  Oexp} give the same result. It would similarly give the same results had we symmetrized the wave function instead of antisymmetrizing. 

This result means that if $\phi_1$ and $\phi_2$ have no spatial overlap then one cannot determine if the wave function has been symmetrized, antisymmetrized, or has not been (anti)symmetrized at all. 

This argument generalizes simply to the many-particle case. If we have two many-body wave functions where there is no spatial overlap between the two, then it is not possible to tell if we have symmetrized, antisymmetrized, or not (anti)symmetrized at all.

\section{Clustering Diagnostic In Detail}
\label{app: cluster}

It is well known theoretically that in high magnetic fields, electrons can form stripes or clusters in partially filled Landau levels~\cite{Moessner, Fogler, Koulakov, Fogler2001, Goerbig04a, Knoester16, Dora23}. Quite a few experiments seem to support this picture as well~\cite{Lilly1, Du1, Friess_2018, Bennaceur, Pollanen, Pan, Msall, Wang, Friess2017, Shingla2023, Cooper, Eisenstein2002, Fu, Do}. While much of the theoretical understanding of these phases comes from simple Hartree-Fock analysis, there is ample reason to believe that this may not always be a complete description of the situation. For example, in Refs.~\cite{LeeJain, Goerbig}  phases of matter were predicted where {\it composite fermions} form clusters --- thus going far beyond Hartree-Fock analysis. One might imagine a situation where one has a wave function and then needs to diagnose whether clustering is occurring. If the clusters are pinned by disorder, one would see density modulation. However, without any disorder, the clusters may be free to move, and the clustering may be harder to diagnose. While the clustering information is certainly contained in particle correlation functions, this information may be hard to extract from typically measured diagnostics such as the pair correlation function.  This is the motivation for constructing a new cluster diagnostic.

\subsection{Careful Definition}

The idea of this diagnostic is to determine if a finite cluster of particles tends to stick together and then the clusters repel. The tool we use is to look at the distance from one particle to the $k^{\rm th}$ closest particle to that one particle.  

Given a configuration of particles ${\bf r}_1, {\ldots}, {\bf r}_N$, let us define the average distance to the $k^{\rm th}$ closest particle in the following way: 

Let $r_{ij} {=} |{\bf r}_i {-} {\bf r}_j|$ be the distance from particle $i$ to particle $j$  (with $i,j {\in} 1{\ldots} N)$. Fixing the index $i$, sort the values of $r_{ij}$ for $j{\neq} i$ in order of smallest to largest distances. The $k^{\rm th}$ value in this list we define to be $d_i(k)$ for $k {=} 1, {\ldots}, N{-}1$. i.e., 
\begin{equation}
d_i(k) = k^{\rm th} \mbox{ smallest value of $r_{ij}$ for fixed $i$ over $j{\neq}i$. }    
\end{equation}
Averaging over $i$, we define $\bar d(k)$ to be the mean of $d_i(k)$ over all values of $i$.  Finally, we average over particle configurations in a normalized wave function $\Psi$ to define 
\begin{equation}
\label{eq:  def_Rk}
    R(k)  = \langle  \Psi | \bar d(k)  |\Psi \rangle .
\end{equation}
Given that the area within a distance $r$ is $\pi r^2$, for a liquid state with uniform density $\rho$, roughly we expect to find $\pi r^2 \rho$ particles in a disk of radius $r$. We then expect that the $k^{\rm th}$ closest particle to a given particle should be roughly a distance $R(k)$ away with $\pi [R(k)]^2 \rho {=} k$. Thus $R(k)$ should be roughly proportional to $\sqrt{k}$. This is not precisely correct as there could be short-range correlations in a liquid state such as a FQH liquid but it should be correct at large $r$ where these correlations vanish. 

However, in a cluster state, where particles cluster together in groups of size $m$, we should find that instead of $R(k)$ being a smooth function, it should have a clear jump between the $R(m{-}1)$ and $R(m)$, showing that the first $m$ particles are closer together than for a structureless liquid, but then the distance to the next particle (i.e.,  to another cluster) is larger. 

Once we have measured $R(k)$, to highlight jumps in an otherwise smooth function, we construct the following function
\begin{equation}
\label{eq:  def_Gk}
 G(k) = [R(k+1)]^2 - [R(k)]^2,~~~~k=1,2,\cdots, N-2.
\end{equation}
Since we expect $R(k)^2$ to be roughly linear in $k$, i.e., $R(k)^2 {\approx} k/(\pi \rho)$ we then expect $G(k) {\approx} 1/(\pi \rho)$. However, for $m$-clustering behavior (cluster of $m$ particles), a jump in $R(k)$ will show a clear spike in $G(k)$ at $k{=}m{-}1$ (and at other values that are one lower than a multiple of $m$). We will call the quantity $G(k)$ the {\it cluster diagnostic}.

The argument we have given for the rough scaling of $R(k)$ assumes a system on a planar geometry. In practice, small system numerics often use a closed spherical geometry~\cite{Haldane830}, and one might worry that the starting point of our calculation, Area${=}\pi r^2$, might fail. In fact, with a very small bit of geometry (see Appendix \ref{app: geometry}), we can show that, so long as we use the chord distance for $r$, the same formula for the area holds and our calculation remains valid. Thus $G(k)$ should still be approximately constant $G(k) {\approx} 1/(\pi \rho)$ for a liquid state. For a sphere with $N$ particles and unit radius, we then have $G {\approx} 4/N$. However, if we have a monopole of flux $N_\phi$, it is more natural to express density in terms of the magnetic length $\ell_B {=} \sqrt{2 / N_\phi}$ so the expected value of $G$ in a liquid state of $N$ electrons should be  
\begin{equation}
\label{eq: Gexpected}
G \approx  2 N_{\phi} /N.   
\end{equation}

\vspace*{10pt}

\subsection{Numerical Results}

\label{app: clusterexamples}

\begin{figure*}
    \centering
    \includegraphics[width=0.66\columnwidth]{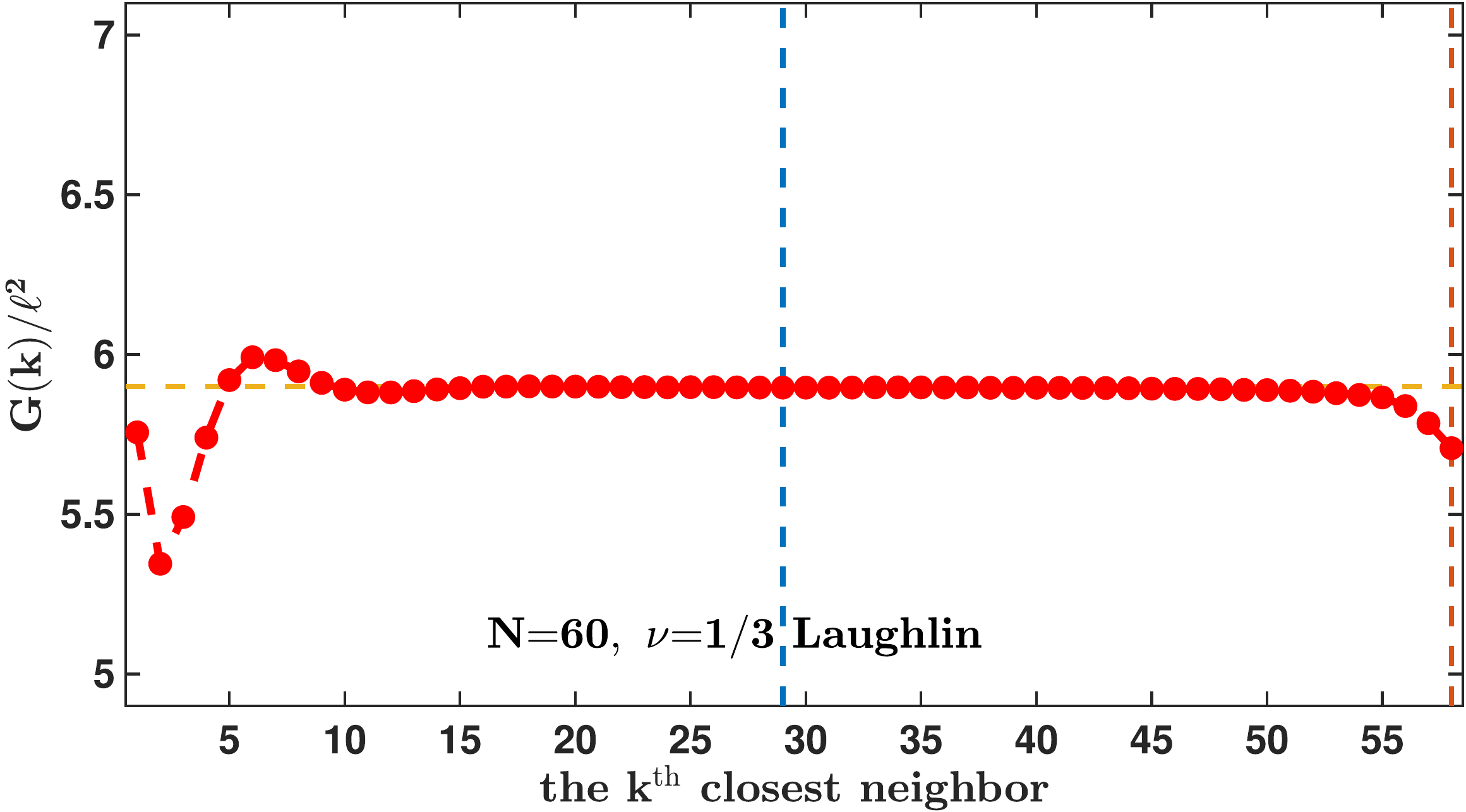}
    \includegraphics[width=0.66\columnwidth]{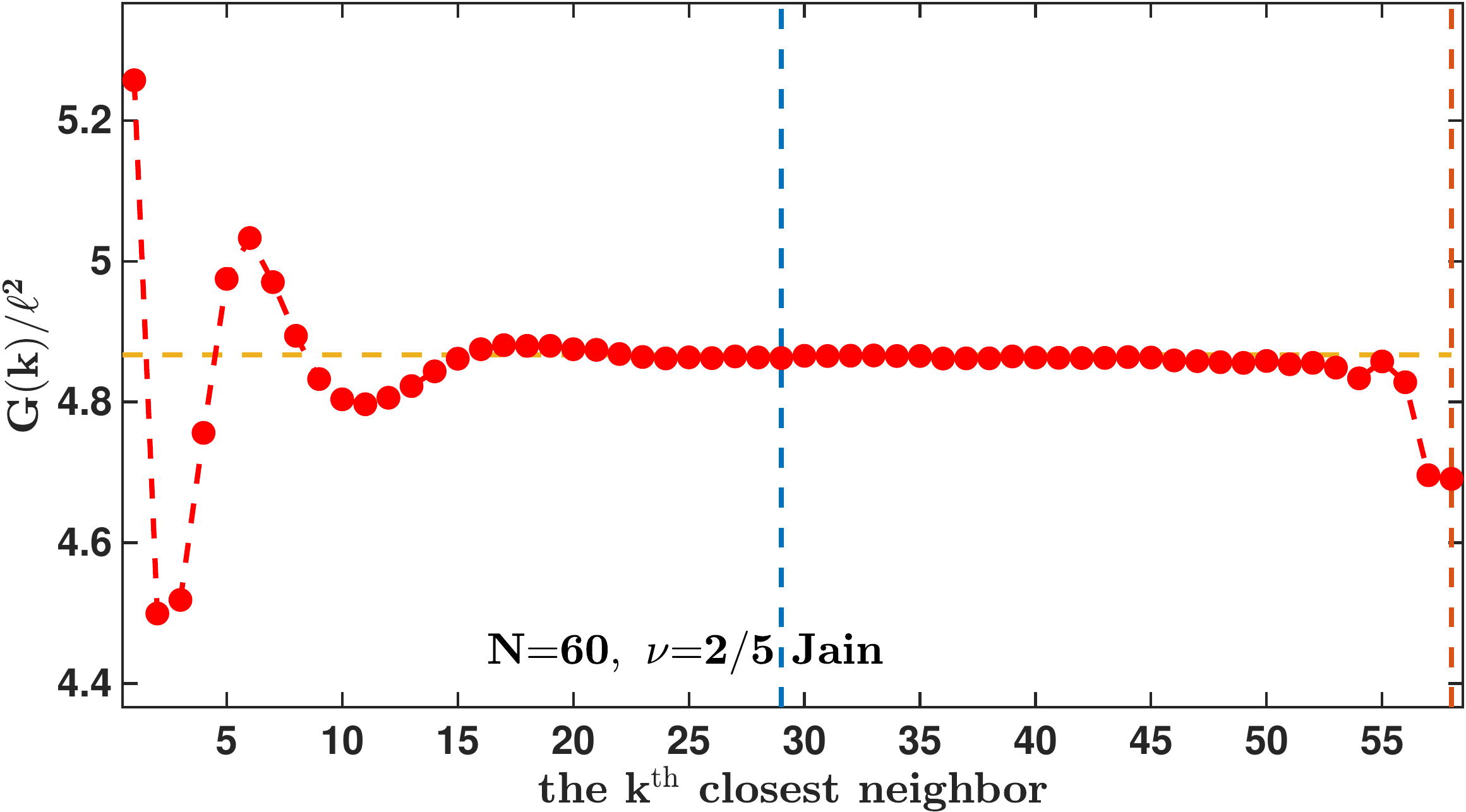}
    \includegraphics[width=0.66\columnwidth]{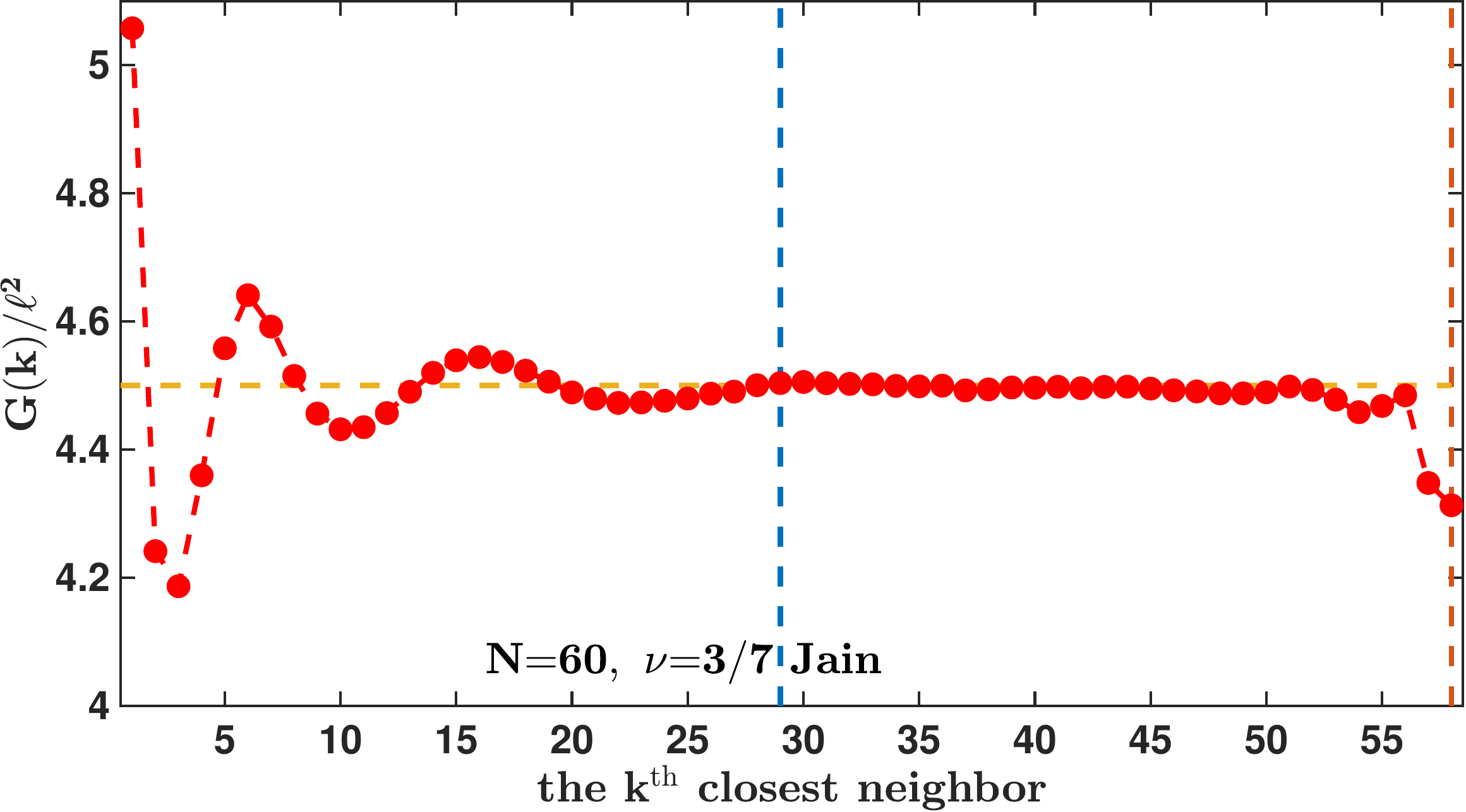} \\
    \includegraphics[width=0.66\columnwidth]{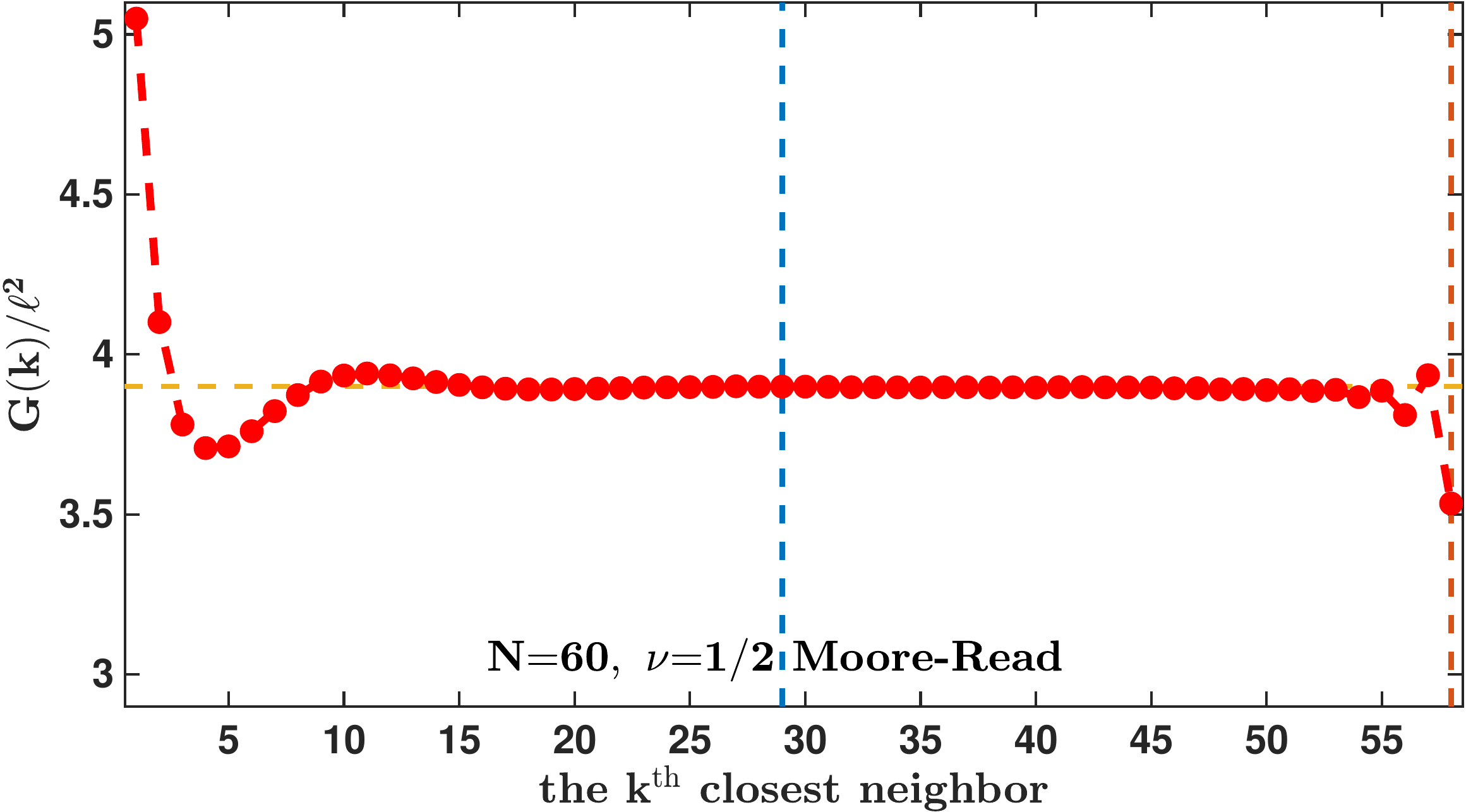}
    \includegraphics[width=0.66\columnwidth]{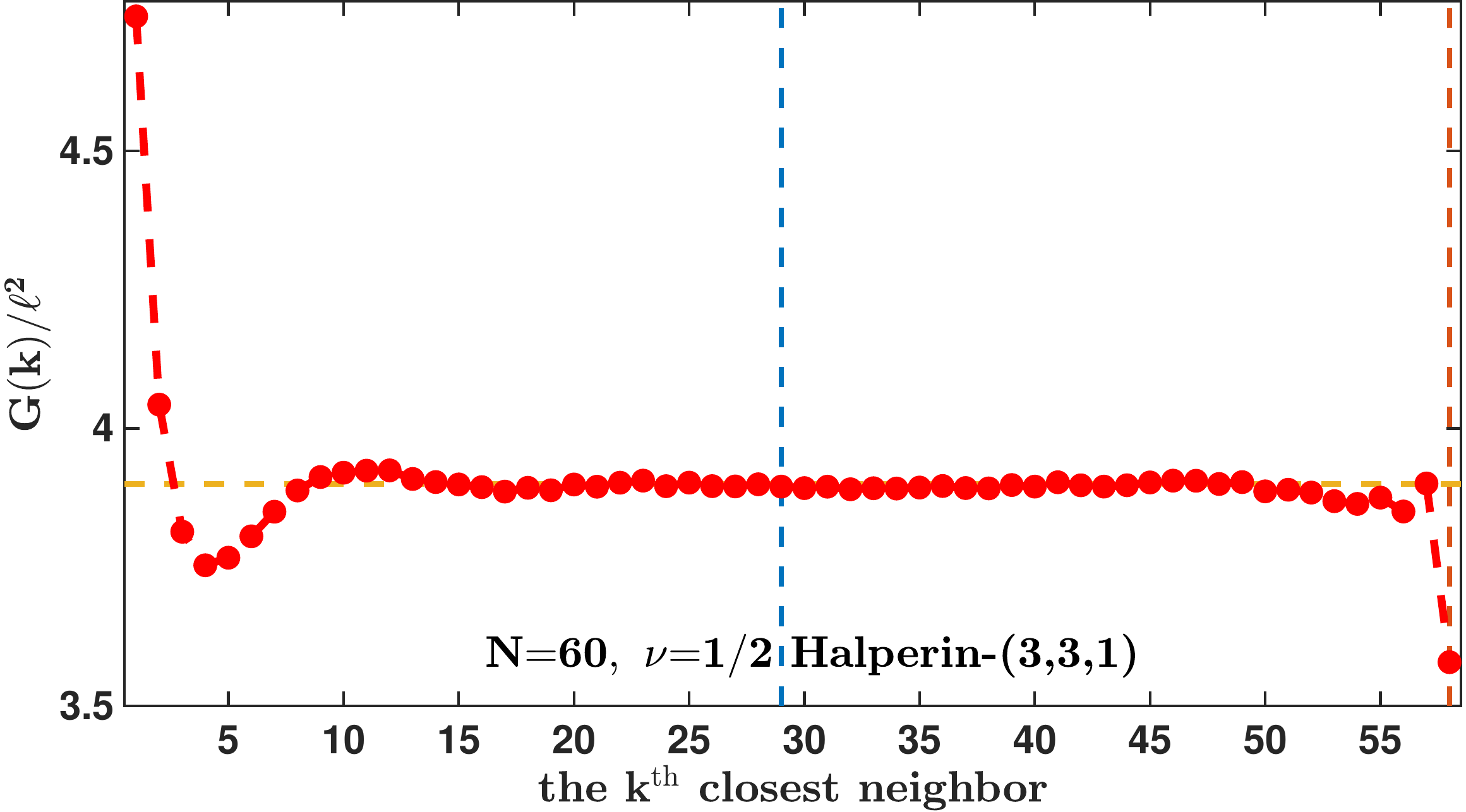}
    \includegraphics[width=0.66\columnwidth]{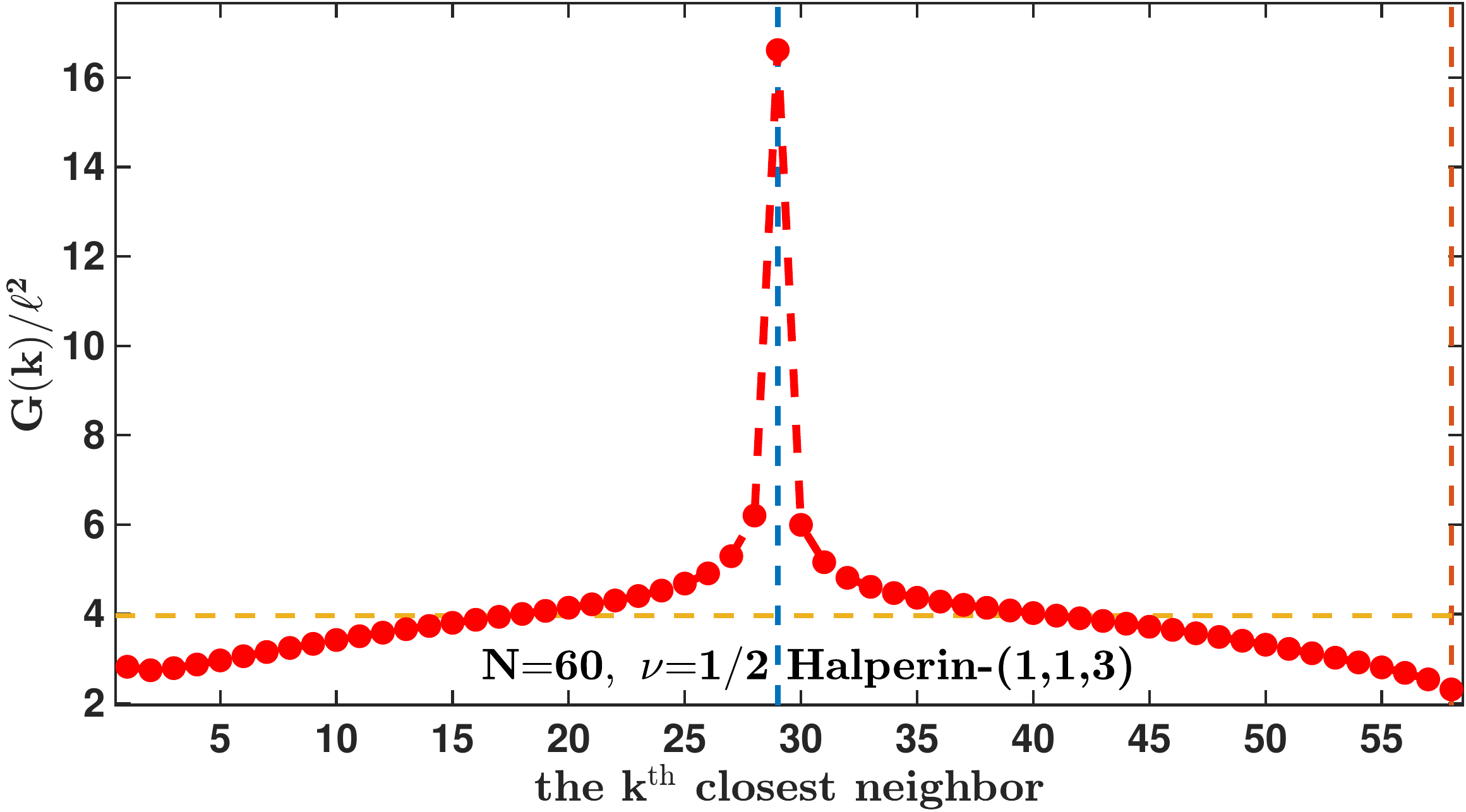}
    \caption{Clustering diagnostic $G(k)$ for well-known fermionic FQH liquid wave functions for $N{=}60$ electrons on the sphere. Top-left: $\nu{=}1/3$ Laughlin, top-center $\nu{=}2/5$ Jain, top-right: $\nu{=}3/7$ Jain, bottom-left $\nu{=}1/2$ Moore-Read and bottom-center $\nu{=}1/2$ Halperin-(3,3,1). Note the scale on the $y{-}$axis. These curves are close to constant and lie very close to the expected value of $G{=}2 N_\phi/N$ which is drawn as the yellow dashed line. For comparison, in the bottom-right panel, we show $G(k)$ for the $\nu{=}1/2$ Halperin-(1,1,3) wave function that is known to describe a state that phase-separates, and here a prominent peak at $k{=}N/2{-}1$ is seen, as expected.}
    \label{fig: known_trial_wfs}
\end{figure*}

We now test this method on several quantum Hall wave functions. In the main text we show the Moore-Read state, as an example of a typical FQH liquid, and the $\mathbb{A}[113]$ state as an example of an antipodally clustered state. As expected, the Moore-Read $G(k)$ is smooth and close to constant, indicating a liquid, whereas the  $\mathbb{A}[113]$ case shows a huge peak at $G(k{=}N/2{-}1)$ indicating separation into two antipodal clusters of $N/2$. 

Here we show some additional data for completeness. In Fig.~\ref{fig: known_trial_wfs} we show $G(k)$ for the 1/3 Laughlin, 1/2 Moore-Read, and Halperin-331 wave functions for $N{=}60$ electrons. Using the Jain-Kamilla approach~\cite{Jain97b0} it is easy to generate composite fermion FQH wave functions~\cite{Jain89} for very large systems. In Fig.~\ref{fig: known_trial_wfs} we also show $G(k)$ for $\nu{=}2/5$ and $\nu{=}3/7$ composite fermion wave functions for $N{=}60$ electrons. In all these cases, the curve is very flat and is close to the expected $G(k){=}2 N_\phi/N$. To contrast, in the bottom-right panel of Fig.~\ref{fig: known_trial_wfs} we also show $G(k)$ for the Halperin-113 wave function which is known to describe a state that phase-separates. As anticipated, the Halperin-113 wave function exhibits a prominent peak in $G(k)$ at $k{=}N/2{-}1$.

\begin{figure}
    \centering
    \includegraphics[width=\linewidth]{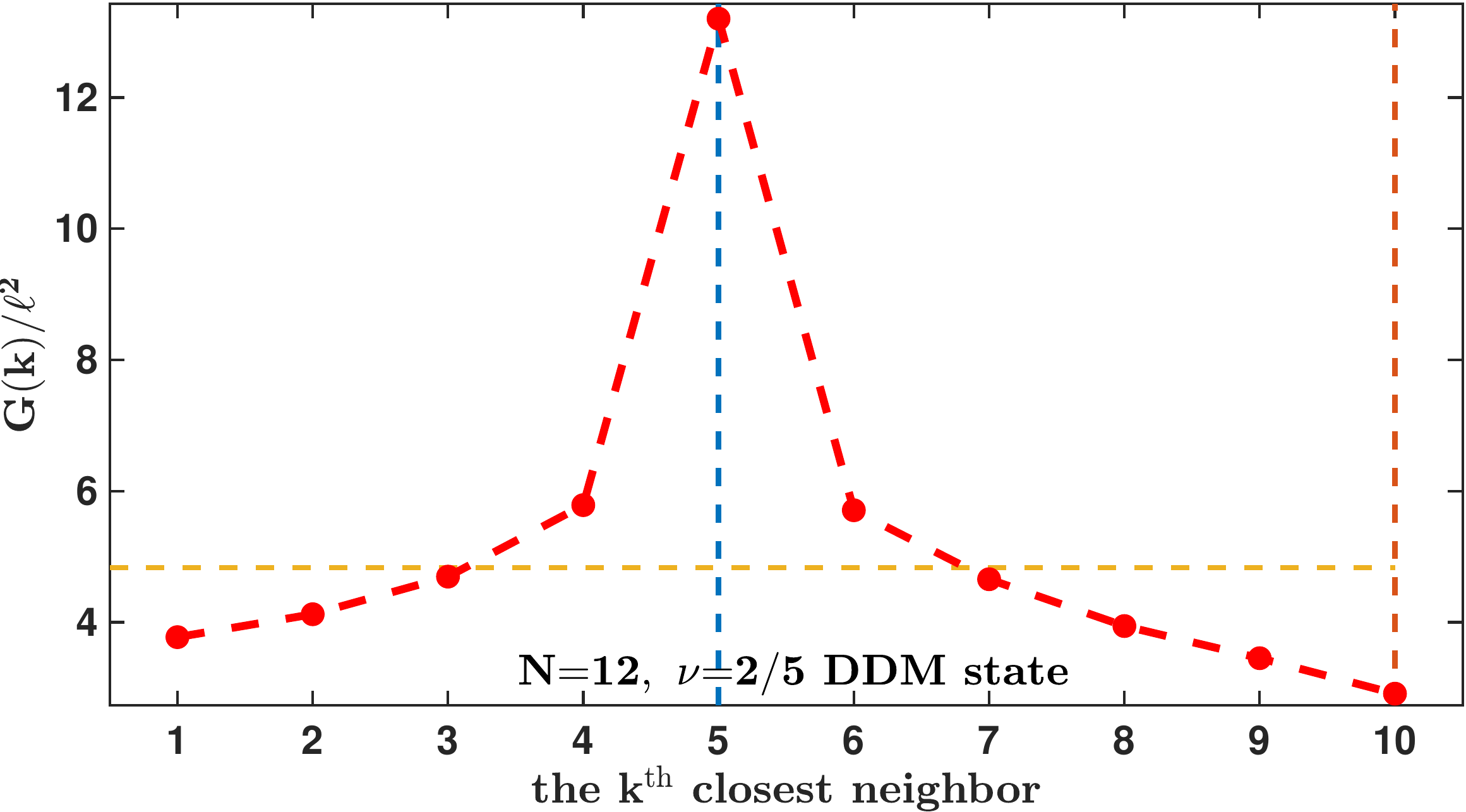}
    \caption{Cluster diagnostic $G(k)$ for the DDM wave function $\Psi(2,3)$ at $\nu{=}2/5$. This clearly shows that the system has phase separated into two antipodal caps.}
    \label{fig:psi23}
\end{figure}

In section \ref{sub: generalwf} we described the wave functions $\Psi(n,m)$ defined by DDM in Ref.~\cite{SSS2}. As an example, in Fig.~\ref{fig:psi23} we show the cluster diagnostic for $\Psi(2,3)$ which is meant to describe $\nu{=}2/5$. As predicted, $G(k)$ shows the large peak at $k{=}N/2{-}1$ indicating that the system has phase separated into two antipodal caps. 

\begin{figure}[h]
    \centering
    \includegraphics[width=\linewidth]{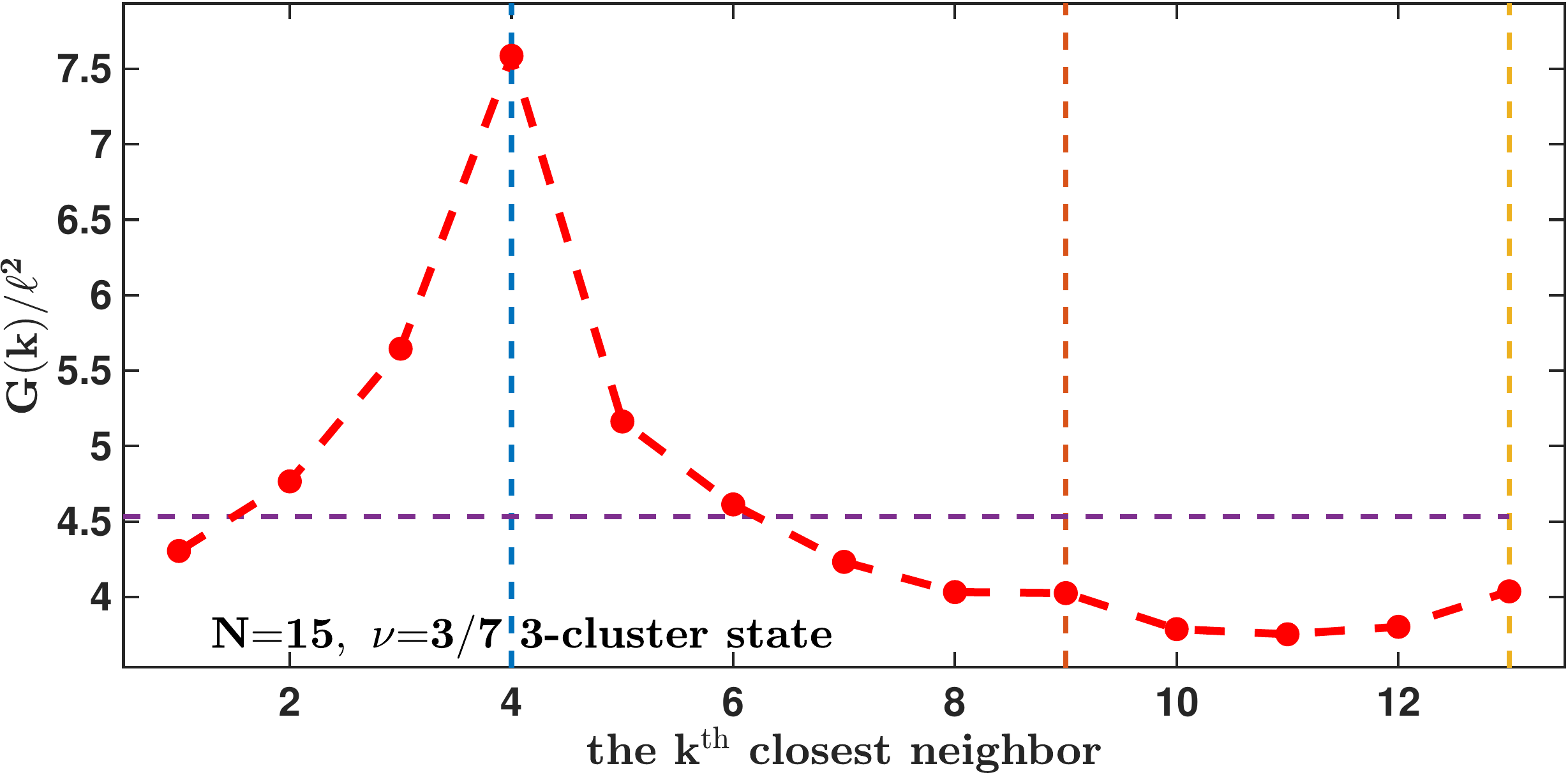}
    \caption{Cluster diagnostic $G(k)$ for the wave function $\Psi_3$ in Eq.~\eqref{eq: 3cluster}. The peak at $k{=}N/3 {-}1$ shows that the system has phase separated into clusters of size $N/3$.}
    \label{fig:3cluster}
\end{figure}

To show that this diagnostic is more generally useful, we consider a wave function that is designed to break into more than two clusters.    Consider dividing particles into three equal groups, $A, B, C$, and writing the wave function using the notation of section \ref{sec: jastrowsection}, we write
\begin{equation}
 \Psi_3 = {\mathbb{A}}[ J  J_{AB}^2 J_{AC}^2 J_{BC}^2 ]  \label{eq: 3cluster}
   \end{equation}
The corresponding cluster diagnostic $G(k)$ is shown in Fig.~\ref{fig:3cluster}. Here the peak is at $k{=}N/3{-}1$ showing that the system has phase separated into three clusters. For this small system, it is only barely possible to discern a small peak at $k{=}2 N/3 {-}1 $ as well. 

\

\subsection{Geometry of the Sphere}
\label{app: geometry}

We consider a sphere of unit radius. The chord distance from one point to another point is $r_{c}{=}2 \sin(\theta/2)$ where $\theta$ is the angle between the two points. The area enclosed within an angle $\theta$ is 
$$
\mbox{Area}  = 2 \pi \int_0^\theta d\theta' \sin(\theta') = 4 \pi \sin^2(\theta/2) = \pi r_{c}^2
$$
Thus the expression for the area enclosed within a (chord distance) radius $r_{c}$ is still $\pi r_{c}^2$.

\section{Theorem About Flatness of Branches of Entanglement Spectrum at Large and Small $L_z$}

\label{app: theorem}

{\bf Theorem:} Consider any $L{=}0$ eigenstate for an even number $N$ of electrons in a Landau level on a sphere with flux $N_\phi$ odd. If we partition the system at the equator with $N/2$ electrons on each side, the largest possible angular momentum that can occur in the orbital entanglement spectrum is $L_z^{\rm max}{=}(N/8)[2 (N_\phi{+}1){-}N]$ [see Eq.~\eqref{eq: Lzmax}]. Defining $L_z^{\rm crit}{=}L_z^{\rm max}{-}(N_\phi{-}N{+}1)/2$, if there exists a mode in the orbital entanglement spectrum having $L_z^0{\geq}L_z^{\rm crit}$ with entanglement energy $\xi^0$, then there also exist modes with entanglement energy $\xi^0$ for all $ L_z^{\rm crit}{\leq}L_z{\leq}L_z^0$.  

Similarly, there is $L_z^{\rm min}{=}N^2/8$ which is the smallest possible angular momentum and $L_z^{\rm crit'}{=}L_z^{\rm min}{+}N/2$. If there exist modes in the orbital entanglement spectrum having $L_z^1{\leq}L_z^{\rm crit'}$ with entanglement energy $\xi^1$, then there also exists modes with entanglement energy $\xi^1$ for all $ L_z^{\rm crit'}{\geq}L_z{\geq}L_z^1$.  

Further, we note that the total count of the number of modes starting from $L_z{=}L^{\rm max}_z$ and counting down can be no greater than the usual bosonic edge mode counting $1,1,2,3,5,7,11,{\ldots}$ for $L^{\rm max}_z{\geq}L_z{\geq}L^{\rm crit}_z$ (i.e., the counting of integer partitions). Similarly, the counting of modes starting from $L_z^{\rm min}$ and counting up can be no greater than the usual bosonic edge mode counting.

\vspace*{10pt}

A more succinct way to state this theorem is that all branches of the orbital entanglement spectrum are exactly flat for $L_z {\geq} L_z^{\rm crit}$ and for $L_z {\leq} L_z^{\rm crit'}$, and the mode counting coming down from $L_z^{\rm max}$ or going up from $L_z^{\rm min}$ is no greater than the counting for a bosonic edge. 

We note that this principle is seen in many entanglement spectra measured previously~\cite{RegnaultReview} (particularly nice examples are shown,  for example, in the Coulomb spectra of  Ref.~\cite{Sterdyniak_2011}), although, to our knowledge, it has not been commented on previously.

We will start by looking at the case for $L_z^{\rm max}{\geq}L_z{\geq}L_z^{\rm crit}$. To prove the theorem we start by considering the Fock states that can give a particular $L_z$ in the entanglement spectrum. Considering only the northern hemisphere, we need to construct a state with angular momentum $L_z$ and the dimension of the corresponding Hilbert space is ${\rm dim}(L_z)$ which has the values $1,1,2,3,5,7,11,{\ldots}$ for angular momentum $L^{\rm max}_z, L^{\rm max}_z{-}1,{\ldots}$. Here we can think of the $L_z^{\rm max}$ state as being the $\nu{=}1$ (completely filled) droplet at the north pole and then we are examining the edge excitations of this droplet. At $L_z{<}L^{\rm crit}_z$ one of the edge modes hits the equator, so this counting is no longer valid. To calculate $L_z^{\rm crit}$ we note that in the $L_z^{\rm max}$ state we fill all orbitals with $L_z {\geq}(N_\phi{-}N{+}1)/2$. If we lower this smallest filled angular momentum $[(N_\phi{-}N{+}1)/2{+}1]$ times it crosses through zero. Thus $L_z^{\rm crit}$ is $L_z^{\rm max}{-} (N_\phi {-} N {+} 1)/2$. The counting of ${\rm dim}(L_z)$ immediately implies the maximum number of modes we can have in the entanglement spectrum as given in the second paragraph of the theorem.

We now think about sewing together the two subgroups of $N/2$ particles to obtain an $L{=}0$ state. It is worth recalling an important property of Clebsch-Gordon coefficients~\cite{Edmonds}
\begin{equation}
\label{eq: clebsch}
 \langle j_1 m_1 ;  j_2 m_2 | 0 0 \rangle = \delta_{j_1, j_2} 
\delta_{m_1, -m_2}  \frac{(-1)^{j_1 - m_1}}{\sqrt{2 j_1 + 1}}.
   \end{equation}
This formula means that to construct an $L{=}0$ state of the entire system we must sum all $m_1{=}{-}m_2$ values up to $j_1{=}j_2$, with {\it equal amplitude} of the different possible $m$ states. This will imply that the entanglement weight of modes must remain constant as a function of $m$ (in this case $L_z$) until the electrons cross the equator at $L_z{<}L_z^{\rm crit}$.

Let us be a bit more precise here. To obtain an entanglement spectrum we make a Schmidt decomposition of our wave function, we partition their orbitals into two halves, (here $A$ and $B$ will be the north and south hemispheres respectively)
\begin{equation}
    |\Psi\rangle = \sum_i e^{-\xi_i/2}  |A,i\rangle \otimes |B, i\rangle,
\end{equation} 
and (assuming $\Psi$ has $L{=}0$) we plot the entanglement energies $\xi_i$ as a function of the angular momentum $L_z^A$ of the $A$ part given that $A$ and $B$ both contain $N/2$ particles. Let us divide this wave function into pieces depending on whether $L_z$ is less than $L_z^{\rm crit}$. 
$$
\sum_{i, L_z^A \geq L_z^{\rm crit}}  e^{-\xi_i/2}  |A,i\rangle \otimes |B, i\rangle +  \sum_{i, L_z^A <  L_z^{\rm crit}} e^{-\xi_i/2}  |A,i\rangle \otimes |B, i\rangle
$$
If we are interested in the entanglement spectrum for $L_z{\geq}L_z^{\rm crit}$ we can ignore the second sum entirely. Since $|\Psi\rangle$ has $L{=}0$ it must be annihilated by both $L_+$ and $L_-$. For states $|A,i\rangle$ and $|B, i \rangle$ which have $|L_z|{>}L_z^{\rm crit}$ we can raise or lower these states, and all of the particles in $|A,i\rangle$ will remain in the north hemisphere and all of the particles in $|B,i\rangle$ will remain in the south hemisphere. Thus the $L{=}0$ condition becomes (using $L_{\pm}{=}L_{\pm}^A{+}L_{\pm}^B$)
$$
0 = \sum_{i, L_z^A \geq L_z^{\rm crit}}  e^{-\xi_i/2}  (L_\pm^A |A,i\rangle \otimes |B, i\rangle +  |A,i\rangle \otimes L_\pm^B|B, i\rangle.
$$
Let us consider the case of $L_-$. Suppose some $|A,i\rangle$ is a highest weight state, i.e., $L_z^A {=} L^A {=} j {>} L_z^{\rm crit}$, which we write as $|j,j\rangle_A$. Let the corresponding $|B,i\rangle$ be $|j', {-}j\rangle$ (with $j'$ not necessarily equal to $j$ yet). The only way we can arrange for this expression to vanish is when $j{=}j'$ and if the sums are of the form   
 \begin{equation}
 \label{eq: Lsum}
\propto \left[ \sum_{p=0}^{p^{\rm crit}} (-1)^p |j, j-p\rangle_A \otimes |j, -j+p\rangle_B  \right] + \ldots
      \end{equation}
which we can check vanishes term by term when $L_-$ or $L_+$ is applied  (i.e., we are reproducing Eq.~\eqref{eq: clebsch} term by term). Here, $p^{\rm crit}$ is chosen to be the value for $L_z {=} L_z^{\rm crit}$. We can similarly run this argument for any $|j, m\rangle$ with $m {\geq} L_z^{\rm crit}$ by applying $L_+$ until we reach a highest weight state. We have thus arranged for $L_{\pm}$ to vanish when applied to our sum, so long as we have $L_z {\geq} L_z^{\rm crit}$. The ${\ldots}$ in Eq.~\eqref{eq: Lsum} indicate states with $L_z{<}L_z^{\rm crit}$ for which electrons have crossed over the equator, and we cannot obtain a corresponding $|j, m\rangle$ by just applying $L_{\pm}$ to one hemisphere. However, we are not interested in these as long as we restrict our attention to $L_z {\geq} L_z^{\rm crit}$. Because of the structure of the sum in Eq.~\eqref{eq: Lsum} required to obtain an overall $L{=}0$ state, we see that any entanglement mode with $L_z {>} L_z^{\rm crit}$ must come in a multiplet of equal entanglement energy stretching from some $j$ (the $L$ angular momentum of the $A$ wave function) down to $L_z^{\rm crit}$. 

We note that this theorem might easily be generalized to account for $N_\phi$ being even, or unequal partitions, either with $N_A \neq N_B$ or even splitting the sphere at a longitude which is not the equator. 

\subsection{Particle-Hole Conjugation}

The above argument is based on the idea that the highest $L_z$ state is the cluster of electrons at the north pole. It is also possible to run the same arguments using the fact that the {\it lowest} $L_z$ state is the cluster of {\it holes} at the north pole. To do this, we note that the angular momentum of a filled hemisphere is $L_z^{\rm filled}{=}(N_\phi{+}1)^2/8$. We then calculate $L_z^{\rm min}$ as $L_z^{\rm filled}{-}L_z^{\rm max}(N {\to} N_\phi{+}1{-}N)$ as defined in Eq.~\eqref{eq: Lzmax} which gives
$$
L_z^{\rm min} = \frac{1}{8} N^2.
$$
Correspondingly there is a second $L_z^{\rm crit}$ for holes given by $L_z^{\rm crit'}{=}L_z^{\rm filled}(N{\to}N_\phi{+}1{-}N){-}L_z^{\rm crit}(N {\to}N_\phi{+}1{-}N)$ which is given by
$$
L_z^{\rm crit'} = L_z^{\rm min} + N/2.
$$

\subsection{Regarding Fig.~2a of Ref.~\cite{SSSReply}}
\label{sub: error}

In all of the orbital entanglement spectra (zero temperature, $L{=}0$, $N/2{+}N/2$ bipartition, and $N_\phi$ odd) we have examined, we have found the above theorem to hold with one exception. The data shown in Fig. 2a of DDM's reply Ref.~\cite{SSSReply} does not appear to satisfy this theorem. While $L_z^{\rm max}$ appears to be correct, no mode extends down to $L_z^{\rm crit}$. We conclude that there must have been some error in creating this data.

\bibliography{bibfile}

\end{document}